# A Drone-based Networked System and Methods for Combating Coronavirus Disease (COVID-19) Pandemic

Adarsh Kumar[1], Kriti Sharma[1], Harvinder Singh[2], Sagar Gupta Naugriya[3], Sukhpal Singh Gill[4], and Rajkumar Buyya[5]

*[1]Department of Systemics, School of Computer Science, University of Petroleum and Energy Studies, Dehradun, India*
*[2]Department of Virtualization, School of Computer Science, University of Petroleum and Energy Studies, Dehradun, India*
*[3]Indian Robotics Solution Pvt. Ltd. (IRS), New Delhi, India*
*[4]School of Electronic Engineering and Computer Science, Queen Mary University of London, UK*
*[5]Cloud Computing and Distributed Systems (CLOUDS) Laboratory, School of Computing and Information Systems, The University of Melbourne, Australia*

## Abstract

Coronavirus disease (COVID-19) is an infectious disease caused by a newly discovered coronavirus. It is similar to influenza viruses and raises concerns through alarming levels of spread and severity resulting in an ongoing pandemic worldwide. Within eight months (by August 2020), it infected 24.0 million persons worldwide and over 824 thousand have died. Drones or Unmanned Aerial Vehicles (UAVs) are very helpful in handling the COVID-19 pandemic. This work investigates the drone-based systems, COVID-19 pandemic situations, and proposes an architecture for handling pandemic situations in different scenarios using real-time and simulation-based scenarios. The proposed architecture uses wearable sensors to record the observations in Body Area Networks (BANs) in a push-pull data fetching mechanism. The proposed architecture is found to be useful in remote and highly congested pandemic areas where either the wireless or Internet connectivity is a major issue or chances of COVID-19 spreading are high. It collects and stores the substantial amount of data in a stipulated period and helps to take appropriate action as and when required. In real-time drone-based healthcare system implementation for COVID-19 operations, it is observed that a large area can be covered for sanitization, thermal image collection, and patient identification within a short period (2 KMs within 10 minutes approx.) through aerial route. In the simulation, the same statistics are observed with an addition of collision-resistant strategies working successfully for indoor and outdoor healthcare operations. Further, open challenges are identified and promising research directions are highlighted.

## 1. Introduction

Coronavirus disease (COVID-19) is a transferable illness that has recently been identified [1]. This infection was unfamiliar before the occurrence of the Wuhan chain in December 2019 and within eight months (by August 2020), over 24.0 million persons are infected and over 824,162 have died. COVID-19's most commonly recognized symptoms are fever, tiredness, and dry cough and some of the people suffered from throbbing pain, nasal clog, runny nose, sore throat, or diarrhea [2]. The old people with medical conditions such as hypertension, heart issues, or diabetes are enduring with illness; individuals with fever, cough, and trouble breathing should seek immediate medical care. This virus spreads between people during close contact, i.e. at a minimum distance of one meter (3 feet), through small beads during hacking, sniffling, or talking [2]. These beads are delivered during exhalation, usually falling to the ground or to the surface instead of being contaminations over long distances [3]. The virus can survive upto *72 hours* on most *surfaces.* Recommended protective measures include hand washing, closing the mouth while hacking, keeping away from others, observing, and self-isolating of persons associated with being infected [2]. This led to usage of transportation restrictions, isolations, lockdowns, stay at occupational risk assessments and the closures of facilities. Coronavirus (CoVs) has evolved as a significant global virus since 2002 in different forms affected thousands of people in multiple countries [3].

Drone-based COVID-19 health and respiratory monitoring platforms creation is being explored by the Australian Department of Defence for health monitoring and detection of infectious and respiratory conditions including monitoring temperatures, heart and respiratory rates, amongst crowds, workforces, airlines, cruise ships, potential at-risk groups, i.e., seniors in care facilities, convention centres, border crossings or critical infrastructure facilities. [4]. 'Drone' a term usually used for an air vehicle that flies like other aviation craft (airplane/pilot) but with a difference of pilot. Traditional aircraft are timely operated by pilots (Autopilot mode is no distinct), this is what makes drone different. Under condition when an unmanned aerial vehicle is in aerospace, it is termed as "Platform". When external hardware or embedded systems are implemented to it is termed as "Payload". Attaching payload to platform results in a drone that can be used in various applications with increased efficiency and accuracy. In [4][5], it is found that drones are widely used in the present COVID-19 pandemic. It is used for monitoring, vigilance, thermal scanning, medication, food supply, alter system etc. In their use, data collection centralization and analysis is a major challenge. The features of present drone-based systems can further be enhanced by integrating the features of measuring social distancing, COVID-19 monitoring, and data collection using artificial intelligence (AI), thermal imaging, sanitization with data analytics, record keeping etc. Understanding the necessity and requirements of drone-based system enhancements for the smart healthcare system, ***the main objectives of this work are:***

[1] Corresponding author at: Department of Systemics, School of Computer Science, University of Petroleum and Energy Studies, Dehradun, India. E-mail addresses: adarsh.kumar@ddn.upes.ac.in (A. Kumar), meetkritisharmahere@gmail.com (Kriti Sharma), hsingh@ddn.upes.ac.in (H. Singh), sagar@irs.org.in (Sagar Gupta Naugriya), s.s.gill@qmul.ac.uk (S.S. Gill), rbuyya@unimelb.edu.au (R. Buyya).

- To propose collision-free zone-based single and multi-layered drones movement strategies.
- To propose an artificial intelligence-based system that collects the data through drones, analyze and provides the necessary security measures.
- To propose a multi-layered architecture that collects the information from drones and exchanges with edge, fog and cloud servers for necessary computing, data sharing, and analytics.
- To simulate the drone-based system for COVID-19 operations such as monitoring, control, thermal imaging, sanitization, social distancing, medication, data analytics, and statistics generation for the control room.
- To implement a real-time drone-based system for sanitization, monitoring, vigilance, face recognition, thermal scanning etc. in COVID-19 hotspots.
- To design and display the statistics of the drone-based smart healthcare system in a control room.

This work starts with reviewing the necessity to design, develop, simulate, and implement a drone-based healthcare system for COVID-19 scenario. In consideration of existing drone-based systems and their features, a drone-based system suitable for COVID-19 or other influenza viruses' pandemic situation is proposed. The proposed approach integrates artificial intelligence processes for data collection, analysis, statistical visualization, sharing, and decision-making. In this work, both simulation and real-time implementation are carried out for COVID-19 operations (sanitization, medication, monitoring, thermal imaging, etc.). In the real-time drone-based implementation, a drone is designed, developed and tested for COVID-19 operations in the Delhi/NCR, India with the approval of government authorities. In the simulation, multiple drone scenarios are considered for COVID-19 operations. Further, multiple drones collision-resistant strategies and their COVID-19 operation in outdoor and indoor activities are proposed and experimented for evaluations. We observed that the drone-based approach can cover a wide area in a short duration and it is an effective approach in pandemic situations and indoor patient statistics.

The rest of the paper is organized as follows. Section 2 presents the start-of-the-art over drone-based systems, COVID-19, drone-based movement tracking systems, and usage of drone-based systems in healthcare. Section 3 presents the proposed drone-based architecture for the smart healthcare system using artificial intelligence processes, fog, edge, and cloud computing services. Section 4 presents the collision avoidance algorithms used in drones' network. Section 5 presents the real-time drone-based evaluation scenario along with a simulation approach for proposed drone-based system for various COVID-19 operations. Finally, the paper is concluded in Section 6 along with presenting open challenges and promising new research directions.

## 2. Related Work

Extending the use of a drone from mission-centric, science, or defense sector to social health is of critical importance especially for dealing with COVID-19 epidemic facing the world. This work supports the argument that drones have influenced health quality and relief measures in real life at a significant level. People are living in unprecedented times where the almost whole world has been affected by COVID-19. Worldwide, doctors and medical professionals are working hard to help diagnose patients, nation leaders are suggesting to maintain social distancing, police and health caring units are inspecting areas trying to sensitize the public along with many other measures being taken at all levels [35]. Drones are proving to be of great assistance in all these areas at varying levels. Countries are considering drones to be of great use through various measures [4]-[6]. Few scenarios where drones have effectively escorted society with health supplements are briefly explained as follows.

- In Australia, a drone flight across cities detects if someone has a "doubtful" respiratory pattern or not [4]. Sensors fitted in the drone's record body-temperature, heart pulse rate, respiratory rate, and other abnormalities. These measurements are taken in different areas, especially in overcrowded areas. Here, a network of the camera is used for monitoring and medication, and it is found to be effective and useful as well. The designed technology for epidemic and disaster control is saving lives, helping people working in critical infrastructure, and deployment of this technology over a large domain is in progress.

- China is practicing surveillance with more than 100 drones (The MicroMultiCopter company) over many cities [7]. This measure is considered to be useful to prevent viral transmission by alarming people if the inter-personal distance between individuals becomes less than a "specific" value or if people are walking in public places without a mask (similar practices were also followed in Spain, Kuwait, and UAE) [7]. The majority of the countries are supporting the drone-based approach for sanitization, monitoring, thermal scanning, governance, vigilance etc. because it provides a safe way to help humanity.
- Sanitization is regulated in China by spraying disinfectant over mass (Terra drone) [5]. In this experimentation, medicine delivery in fixed areas, sanitization, monitoring, and analysis is performed. This approach is tested with at least 1000 patients and effective in a real-time environment. The proposed approach is found to be more effective in rural areas where resources are scarce especially medical supplies. Further, the drone-based system helps in teaching the people how to wear mass and stop spreading the virus. Lastly, the potential patients are identified, counted, and analyzed using thermal images and measuring the body temperatures. Thus, the proposed approach is found to be very effective in smart healthcare systems.
- In the United States, a personal medical kit for COVID-19 is being delivered by UAVs to remote locations [7]. Like the other drone-based systems, this system is found to be effective in delivering medical and other necessary supplies. In the US, it is found to be effective in rural areas where corona symptoms are found in patients. Likewise, it is recommended that the use of this technology should be increased to a large scale to overcome the situations and help humanity in every aspect that we can do for saving the lives, providing them with their necessities, and establishing healthy communication with everyone in all aspects.
- In India, states like Delhi, Kerala, and Assam are making announcements during surveillance across cities via drones. Maharashtra is a step ahead as it has generated data analysis reports of the area being covered via drones [8] [9]. In overall observations, government authorities in India have given special permissions to their bureaucrats and police officials to use drone-based technology for vigilance, monitoring, medication, sanitization, data analysis, reporting, and future decision-makings. Here, top officials in the government and senior ranked officials are monitoring and taking control over COVID-19 hotspot positions through various means. Drone-based systems are made an advance to handle individual patients or thermal scanning a large area in short as well. Further, they are used for delivering medicines, food supplies, and other necessary equipment weighing a few kilograms. Thousands of drones are deployed in each state of India for similar actions with government permissions and it is observed that the success rate these drone-based systems and networks is very high.

Table 1 shows the comparative analysis of various existing drone-based approaches used in pandemic or other disasters. Table 2 shows the comparative analysis of state-of-the-art drone-based approaches used in various applications (including healthcare systems).

**Table 1:** Comparative analysis of the proposed approach with an existing drone-based system for pandemic or disaster analysis.

| Authors | Year | System Type | Features | | | | | | |
|---|---|---|---|---|---|---|---|---|---|
| | | | A | B | C | D | E | F | G |
| Lum et al. [10] | 2007 | Drones / UAVs deployed for a surgical robot | × | ✓ | ✓ | × | ✓ | × | × |
| Câmara [11] | 2014 | Drone-based Rescuers and disasters scenario system | ✓ | ✓ | × | × | × | ✓ | × |
| Kimet al. [12] | 2017 | Drone-based healthcare services for patients with chronic disease | × | ✓ | ✓ | × | ✓ | × | × |
| Robert et al. [13] | 2018 | Drone-based system for medical services | ✓ | ✓ | ✓ | × | ✓ | × | × |
| Peng et al. [14] | 2018 | Drone-based vacant parking system | ✓ | ✓ | × | × | × | ✓ | × |
| Pirbhulal et al. [15] | 2019 | Time-domain feature-based medical system using wearable sensor devices | ✓ | ✓ | × | × | ✓ | × | × |
| Ullah et al. [16] | 2019 | Drone-based multi-layered architecture with healthcare use-cases | × | ✓ | ✓ | × | ✓ | × | × |
| Jones et al. [17] | 2019 | Drone-based system in medical drug supply | × | × | ✓ | × | ✓ | × | × |
| Islam and Shin [18] | 2020 | Drone-based system integrated with IoT and blockchain-technology | × | ✓ | × | × | × | ✓ | × |
| Sethuraman et al. [19] | 2020 | Drone-based healthcare system experimented and tested for cyber-attacks. | ✓ | × | × | ✓ | ✓ | × | × |
| Proposed Approach | 2020 | Drone-based system for COVID-19 operations with AI processes | ✓ | ✓ | ✓ | ✓ | ✓ | × | ✓ |

Features: A: Monitoring, B: Data Collection, C: Medication, D: Thermal Imaging, E: Healthcare, F: Other Application, G: AI Trends,

**Table 2:** Comparative analysis of existing drone-based systems for various applications.

<table>
<tr><th>Authors</th><th>Year</th><th>Pros</th><th>Cons</th></tr>
<tr><td colspan="4">Surveys conducted over drone-based systems</td></tr>
<tr><td>Lee and Park [20]</td><td>2017</td><td rowspan="5">Conducted drone-based medical services surveys [13][21].<br>Discussed the drone-based system benefits in rural areas [13][16][21].<br>Discussed drone-based medical and non-medical systems' use-cases [13][16][21].</td><td rowspan="5">Challenges and solutions to congested drone networks are not proposed.<br>All-weather drones and their usage is missing.<br>Surveys over drone-based networks and their implementation issues are not conducted.<br>Discussions over types of laws, procedure, process, and governance is missing.</td></tr>
<tr><td>Robert et al. [13]</td><td>2018</td></tr>
<tr><td>Ullah et al. [16]</td><td>2019</td></tr>
<tr><td>Ullah et al. [21]</td><td>2019</td></tr>
<tr><td>Skorup and Haaland [22]</td><td>2020</td></tr>
</table>

| | | | |
|---|---|---|---|
| | | The pros and cons of drone-based healthcare systems over pandemics (e.g. COVID-19) and disasters are widely discussed [13][21][22]. | |
| **Drone and robot-based system for surgical operations** | | | |
| Harnett et al. [23]<br>Lum et al. [10]<br>Lee and Park [20]<br>Kimet al. [12]<br>Ullah et al. [16] | 2008<br>2007<br>2017<br>2017<br>2019 | Conducted robot and drone-based operations remotely [10][12][23].<br>Measured performances over drone, robot, and control room integration and networks [10][16].<br>Created medical facilities with drone-based operations [10][16][20].<br>Healthcare services for chronic disease in one-to-one connectivity and medical facility is a unique approach for data exchange [12]. | Congested drone-networks construction and operations are not discussed.<br>Drone-collision strategies and unanimous decision-making are not planned.<br>One-to-one connectivity can increase the number of drones' presence in a small region. Thus, it would be very difficult to avoid collisions.<br>Surgical operations with high success rates are yet to be practiced. |
| **Drone-based system for monitoring and data collection** | | | |
| Todd et al. [24]<br>Thiels et al. [25]<br>Ullah et al. [16]<br>Islam and Shin [18]<br>Sethuraman et al. [19] | 2015<br>2015<br>2019<br>2020<br>2020 | Data collection features and related use cases with 5G networks are discussed [16].<br>An IoT and blockchain integration and interoperation are proposed for providing data security, speed, and QoS in medical data collection [18].<br>Indoor drone-based systems are designed for patient monitoring and medications [24]. These drones are very useful for patients affected with influenza, COVID-19 or airborne viruses. | Drone-networks interoperations, data format standardization, data pre-processing, and redundancy removal techniques are not discussed in detail.<br>The IoT and AI lack in measuring the drone-network performances and QoS performances.<br>Indoor patient monitoring drones are restricted to monitoring, data collection, or intimating for medications only. |
| **Drone-based system for data analysis and decision making** | | | |
| Harnett et al. [23]<br>Ullah et al. [16]<br>Islam and Shin [18]<br>Sethuraman et al. [19] | 2008<br>2019<br>2020<br>2020 | The integration of IoT and blockchain technologies helps in decision making for healthcare scenarios at edge computing levels [10][16].<br>A network of drones helps in collecting the data from remote drone locations. Thus, it is found to be helpful in drone network monitoring as well [19]. | The IoT, edge computing and blockchain technology integration could be enhanced with AI processes and practices.<br>The scarcity of resources over drone devices does not help to process large data and fast decision-making. Thus, federated learning should be adopted for quick and reliable decisions. |
| **Drone-based system for data and product sharing** | | | |
| Kimet al. [12]<br>Pirbhulal et al. [15]<br>Ullah et al. [16]<br>Jones et al. [17]<br>Islam and Shin [18]<br>Sethuraman et al. [19] | 2017<br>2019<br>2019<br>2019<br>2020<br>2020 | Time-domain feature-based drone-systems are proposed [15][16].<br>Medical drug supply and order confirmation is proposed [17].<br>Body-area sensor values helped in medical decision-making [18].<br>A network of drones is constituted for cyber-attack detection [19].<br>The low load-carrying drones can cover a large area within a short time (upto 4 kilometres in 20 minutes) [18][26]. | Time-domain feature-based systems do not experiment in real-scenarios.<br>The drone and its network performance analysis are required for optimizing the services while in data or product sharing scenarios.<br>The importance of indoor drone-experimentation is realized in COIVD-19 pandemics. |

***Critical Analysis:*** In the existing literature [27]-[30][37]-[42], various drone-based approaches are proposed for smart healthcare or mission-oriented systems. This approach can be realized from COVID-19 pandemic situations as well. Various challenges and requirements in establishing drone-based smart healthcare design, simulation, implementation, or analysis include: (i) the construction, designing, and analysis of drone-networks for healthcare or other applications are least practiced, (ii) performance and QoS improvement in a resource constraint drone device collecting medical data from heterogeneous sensor, (iii) collision-free drones based network and routing, (iv) multi-layered drone flying, collision detection and resistance strategies, (v) solar-based environment-friendly in-air drone-charging system, (vi) testing of the drone-based system in pandemic situations (such as COVID-19) can be performed for various operations like monitoring, vigilance, sanitization, medication, thermal imaging etc., (vii) applying artificial intelligence in drone-operations, (viii) integration of drone-level federated learning for self-governed processes and analyzing changing COVID-19 viruses, symptoms and strategies, and (ix) address the e-government privacy-preserving concerns that trace the infected and suspected cases nationwide and securely transmit data. It should handle privacy concerns, gives an option to voluntary or mandatory use the application, limits data collection, and usage, handles data destruction diligently, system transparency, and minimal data collection.

## 3. A Drone-based Architecture for Smart Healthcare System

This section proposes a drone-based architecture for the smart healthcare system. In this architecture, Artificial Intelligence trends are integrated as shown in Fig. 1. This integration includes machine learning, and deep learning for data analysis, Internet of Things (IoT) [36], Industrial IoT (IIoT), Internet of Medical Things (IoMT), and Internet of Drones (IoD) for data collection or instruction-based services. AI is applied to design, develop, and make internal processes more efficient. Thereafter, cloud, fog, and edge computing approaches are applied for efficient data storage and processing starting from nearby locations to a distant secure position. Likewise, other technological aspects including commuter movement, profiling, monitoring etc. are recovered, observed, and analyzed for pandemic spreading.

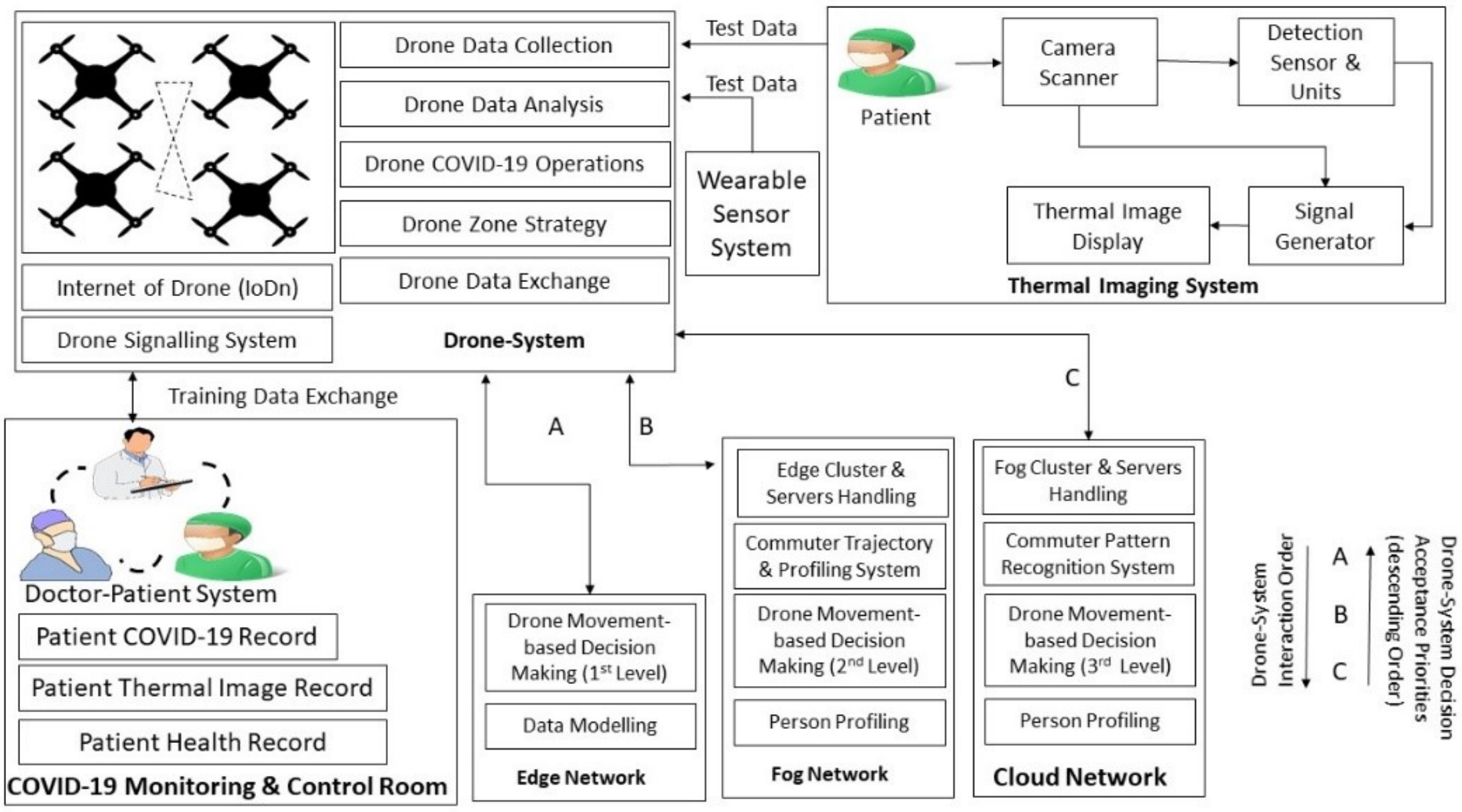

**Figure 1:** Architecture for Drone-based COVID-19 Monitoring, Control, and Analytics in Smart Healthcare System

Fig. 1 shows the proposed architecture for drone-based COVID-19 monitoring, control, and analytics in a smart healthcare system. In this architecture, there are six systems, which are discussed below:

- **Thermal Imaging System:** An alternative to sensor deployment-based test data collection, drone-based cameras can capture the person images and can be useful in social distancing measurements and density-based thermal imaging. Using this system, the camera scanner detects the object and signal for the thermal image display. If the image is not clear for thermal display then it is made clear through detection and unit system.
- **Wearable Sensor System:** In this architecture, it is assumed that sensors are deployed in the observational area. The deployment of the sensor includes wearable sensors, or movement detection sensors at the ground near to target monitoring/serving areas. This helps construct an IoT network using sensors, increase the scale of interconnection and construct IIoT, or IoMT. All these deployments help in collecting the required data, data analysis, and generating statistics. The wearable sensors, movement detection sensors, image processing etc. can be used to monitor the COVID-19 pandemics. Fig. 2 shows a drone-based person monitoring system using wearable sensors. A person under observation is monitored continuously through wearable sensors. Drones placed closer to the population for collecting the patient's data from wearable sensors or thermal imaging receive and store data in drone memory. The stored data is forwarded to big data storage through multiple servers. These servers use edge, fog, and cloud computing for processing, modeling, profiling, and analyze the data. The analyzed and refined data is shared with hospitals with the proper policies and procedures, controlled and governed by the medical board and the federal government. This way, it would be much convenient for both the federal government and hospital to pre-plan the resources in emergency cases. Patient data is securely transferred to the doctor/hospital as and when required. This proposal is found to be very handy to tackle pandemic situations such as COVID-19. In pandemic situations, when data such as time and location of collecting data, population size, person's profiling, methods of collecting data etc. is available then it would be much convenient for everyone to govern the activities and events. This way, it would be much convenient to identify the zero COVID-19 patient and the chain of the pandemic. The integration of wearable sensors and drones are considered in IoT. An IIoT is formulated when multiple stakeholders (e.g. hospital, drug, and equipment supplier, government, medical board etc.) are involved. Further, IoMT is considered to interconnect the medical equipment, drug, patient, and doctor system for clinical trial data fetching and storage. Similar or different clinical trials data is helpful for analysis and advancements.

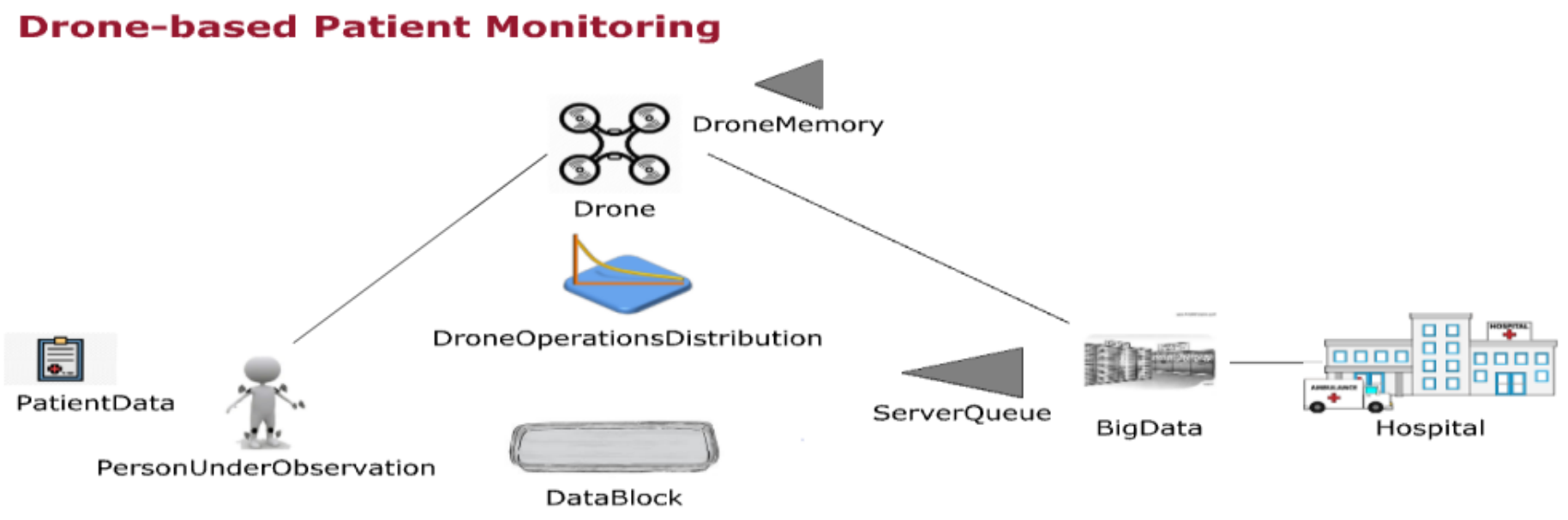

**Figure 2:** Drone-based Person Monitoring System using Wearable-Sensor Data

- **Edge Network and Computing System:** This system applies edge computing for data modeling and initial decision-making. Edge computing does not fetch all drone data. The drone does the self-processing and keeps the data discrete. Edge computing saves time and resources while maintaining the data collection, pre-processing, and analysis in real-time. Thus, edge computing is helpful to drones in making quick real-time decisions. Those initial level decisions are helpful for drones that need instructions and operate within a stipulated time. Here, the major challenge for edge computing is to balance the performance in handling drone-system and fog computing tasks. In the proposed system, edge computing system sometimes needs to make the decision locally and other times it has to send the data to fog servers for further detailed processing. The cost of transferring data to fog servers increases with an increase in the scalability of sensors, IoT, and drone networks. Large data would require resources and intelligence. Here, edge computing shreds its load by performing data aggregation at the initial level and transfer the necessary data to fog or cloud networks as and when required.
- **Fog Network and Computing System:** This system adds fog computing services in the architecture for commuter profiling, monitoring, and decision-making processes executed in the initial phase. Thereafter, data analytics helps in smart and intelligent commuter trajectory profiling, monitoring, and decision-making. In the system, multiple drones collect person's information that differs in attributes. This way of collecting the data is much convenient to make a person's profile. Likewise, a COVID person's profile helps in tracking the COVID-19 cases' chain.
- **Cloud Network and Computing System:** This system applies application-level services for activities such as pattern recognition, monitoring, decision making, and large-scale sanitization are created. High-end cloud computing resources offer capabilities for comprehensive analytics and decision making compared to the other layers.
- **Drone-System:** Drone-based healthcare system is having various advantages over CCTV-based monitoring including (i) it can cover those areas which are hidden in CCTV footage, (ii) drones are multipurpose, it can be used for medicine delivery, sanitization, thermal imaging, scanning, etc., and (iii) drones can monitor the patient from a close position as compared to CCTV. Medical sensors attached to a drone (such as accelerometers, biosensors, MEMS etc.) can measure the patient conditions more closely and accurately compared to CCTV, and (iv) loud noise is not there in every type of drone. There are quite drones (like DJI Mavic Pro Platinum, Parrot Mambo, and DJI Phantom 3 Pro) designs that are more beneficial for indoor hospital systems. In the proposed system, one or more drones move-around and push/pull the required information/instructions from sensors as and when required. The data is processed initially at the drone for initial instructions. Thereafter, it is shared with other systems for further detailed processing. The internet of drones is constituted for longer data transfer and analysis. This internet of drones avoids collisions through either Radar / LiDAR systems or collision avoidance strategies. Apart from Radar/optical systems, collision avoidance strategies are required if a large number of drones are used for different services. Each of these services governs their strategy for drone movement. Using collision avoidance strategies, a pre-planned drone movement strategies could be implemented and short-distance based collisions could further be avoided with Radar/Optical systems.
- **COVID-19 Monitoring & Control Room System:** In this system, drones and area under observations, and their associated statistics are observed. This system helps in monitoring COVID-19 hotspots remotely and plan for necessary actions. Further, individual drone's performance and movement can also be measured and controlled.
- **Information Capturing, Processing and Security Flow:** The proposed drone-based smart healthcare system aims to capture the person's information using drones and then transmit it to the hospital remotely. However, there are data processing, security, and privacy concerns that need to be addressed for successful implementation/considerations. Fig. 3 shows the drone-based information capturing, processing, and security flow in the proposed smart healthcare system. The COVID-19 operation starts with data processing and protection agreement between the Resident Welfare Association (RWA) and the government. In Delhi/NCR region, RWA is a body that represents the interest of people (such as managing facilities, providing safeguards, and organizing events) living in a specified area or society. In this work, RWA appointed **Indian Robotics Solution (IRS)** for COVID-19 operations after an agreement. A team of IRS used a drone system for data collection. The IRS and Local drone data collectors operate the drone and collect the data using a secure tunnel. This secure tunnel uses end-to-end data encryption (using Lightbridge application) for data sharing. Lightbridge application in the drone system ensures that only authenticated devices receive the data. In Lightbridge application, multiple drones in distributed architecture can receive the data in a secure tunnel. Local drone data collector uses cryptographic primitives and protocols for secure data storage. The collected data is shared with the government-appointed medical board either using manual data sharing or automated data sharing process. In the manual data sharing process, a government official collects the data. Whereas, automated data sharing uses end-to-end secure data sharing tunnel. Hospitals collect the data from the government's system using end-to-end secure data sharing tunnel.

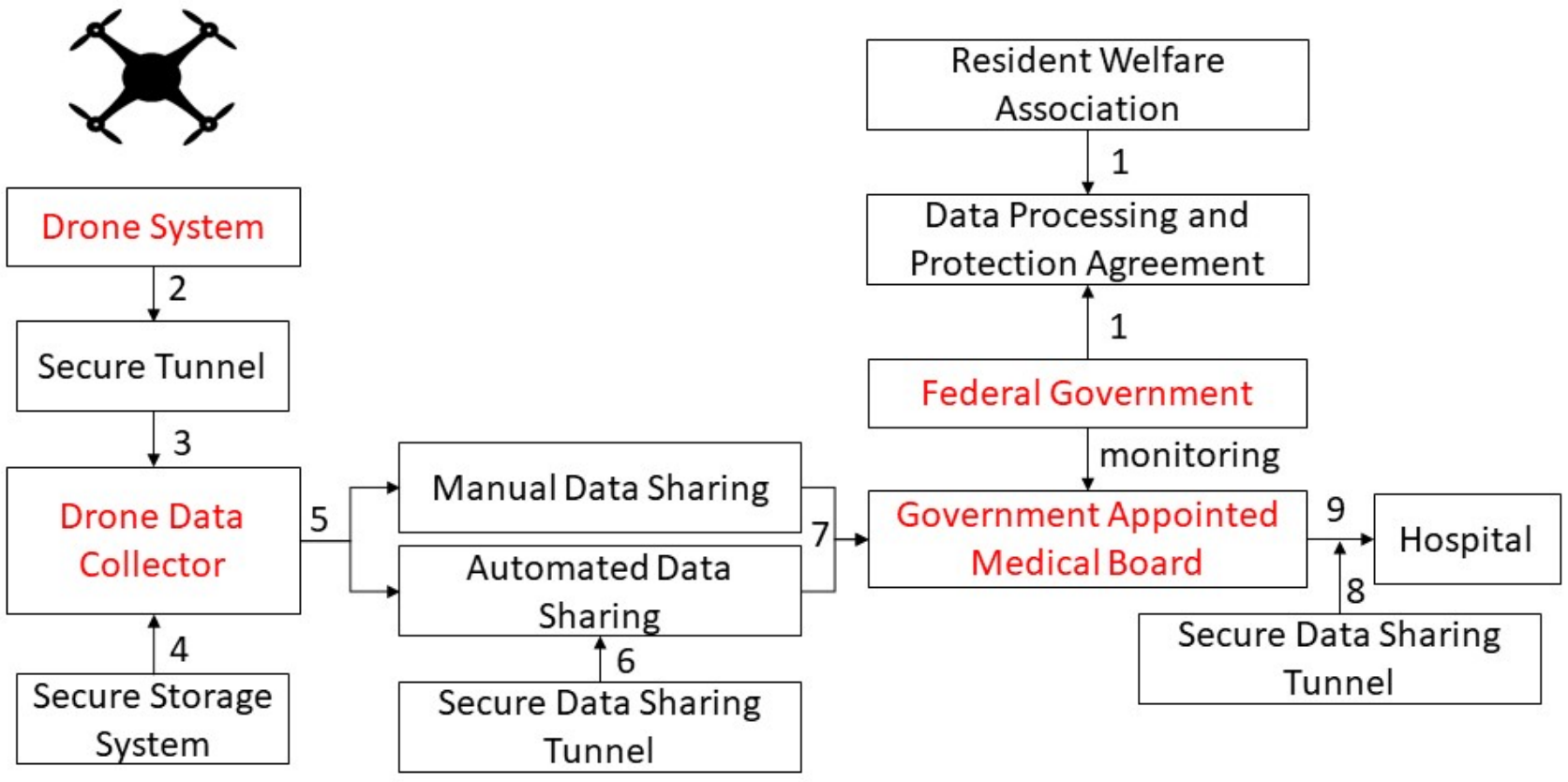

**Figure 3:** Information capturing, processing, and security flow.

## 4. Algorithms and Operational Strategies

AI-based drones capture raw data from IoT networks and can turn them into useful and actionable outcomes. Our approach helps in multiple-drones to collaborate, share, and process the data and trace the pandemic. AI-powered drones are useful in gathering pandemic ground intelligence, assess the COVID-19 virus spreading, thermal imaging for pinpoint, and diagnose issues. All these COVID-19 operation statistics can be efficiently computed if data is collected systemically. Thus, a geographical area is divided into multiple zones for drone-based COVID-19 operations as shown in Fig. 4. Here, square zones of $\tau$ distance are formed and each zone is planned to have an area that can be covered by a single drone. Algorithm 1 shows the drone's federated learning process before sharing its training data with an edge, fog, or cloud network computing. Thus, each drone can monitor its zone based on individual experiences. However, an average value of observations from all zones is collected at the edge computing side for collective decision. Algorithm 1 suggested applying the patient monitoring, thermal scanning, and image identification followed by the patient's body temperature measurement. The scalability of applying a unique solution to drones increases as the decision process moves from edge computing towards cloud computing. In the drone-system (shown in Fig. 1), sensor and IoT-enabled infrastructure can be monitored by drones, and the movements of drones are important to observe for collision avoidance. Thus, the zone-based approach does not allow any collision. In every zone, there is a collision feasibility area. If any drones move to the collision feasibility area then it will send a signal (through LiDAR/RADAR/beam systems) to all neighboring drones for collision avoidance. Now, if any drone wants to transfer its zones due to various reasons such as high-battery drone requirement, longer operation time drone requirement, malfunctioning of certain drones, specialized drones requirement, rotation of drone shifts etc. then zone transfer algorithms with artificial intelligence components based experiences can be applied. The zone-transfer strategies are divided into single or multi-layered algorithms. These algorithms are explained in the following sections.

**Algorithm 1:** Federated learning for drone-based zone operations (monitoring, sanitization, vigilance etc.)

**Goal:** To integrate an individual drone's zone-based self-learning for COVID-19 experiences and securely exchange with an edge, fog, and cloud servers.

**Premises:** Let $N_l$ represents the $l^{th}$ drone-network, $Z_i$ represents the $i^{th}$ zone in the area, $D_j^i$ shows the $j^{th}$ drone in $i^{th}$ zone and $P_k^i$ represents the $k^{th}$ patient in $i^{th}$ zone. $\tau$ indicates the length and width of a single zone. $\delta$ is a time interval of the drone' zone scanning process. $Q_{Z_i}^{N_l}$ shows the QoS measurements for $Z_i$ in $N_l$. $C_{Z_i}^{N_l}$ is the COVID-19 experiences for $D_j^i$ in $Z_i$ of $N_l$. $E^{N_l}$ represents the edge server used for computing $N_l$ statistics. Here, $Z_{i+n+1} = Z_i$.

1. **For each** $N_l$:
2.     **For each** $Z_i$:
3.         Associate $D_j^i$ with each $P_k^i$
4.         **For each** $\delta$ duration:
5.             Collect COVID-19 scanning, thermal image collection, temperature, and other wearable sensor-based measurements
6.                 **If** ($P_k^i$'s body temperature increases with time) **then**
7.                     Start sanitization and medication
8.                 **End If**
9.                 $\delta = \delta + \delta$
10.         **End For**
11.         **Measure** $Q_{Z_i}^{N_l}$ and $C_{Z_i}^{N_l}$
12.     **End For**
13.     **Share** $Q_{Z_i}^{N_l}$ and and $C_{Z_i}^{N_l}$ with $E^{N_l}$
14. **End For**

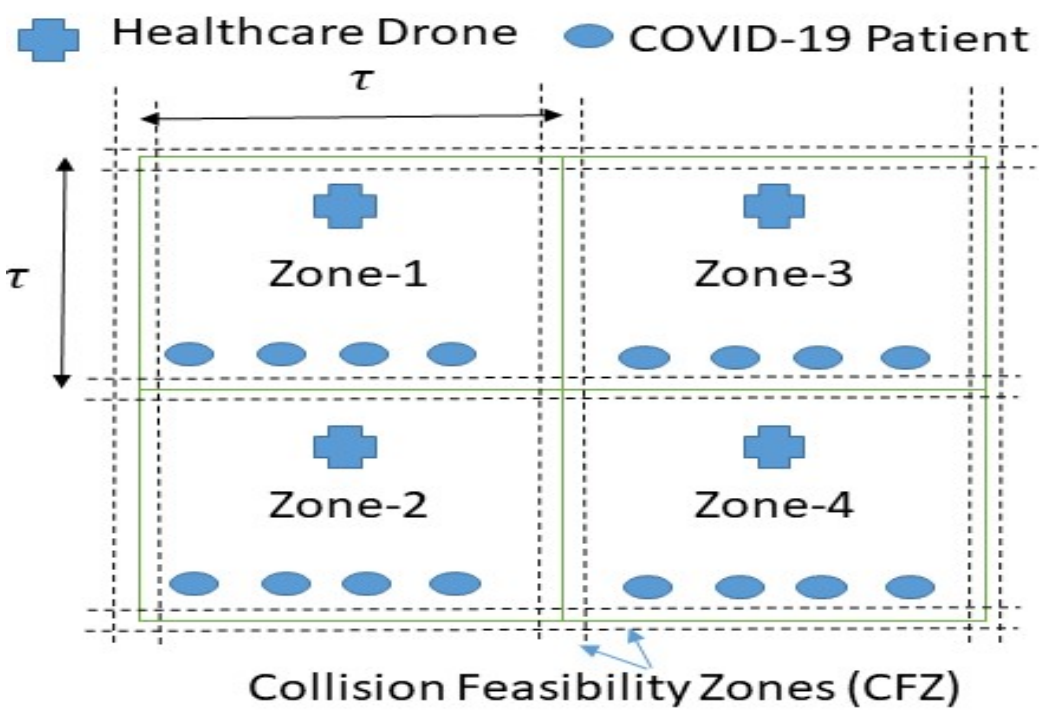

**Figure 4:** Proposed collision avoidance strategy for a drone network.

## *4.1 Single Layer Zone-Transfer Algorithms*

This section presents three zone-transfer algorithms: fixed-area transfer, zigzag transfer, and parallel movements and zone-transfer. Fig. 5 proposes a single-layered fixed-zone transfer strategy. In this proposed approach, drones are allowed to change or exchange their zones using a zone transfer area. This area is pre-planned to have a storage of at least one drone. The exchange is allowed if both the transfer zones (left to right and right to left) linking each other is empty. The details of the proposed single-layered zone transfer strategy are explained in algorithm 2. Fig. 6 shows the simulation of the single-layered zone transfer strategy (in execution). For example, if one drone is present in each zone-1 and zone-3, and they want to exchange their positions then zones transfer areas marked with arrows help them to swap the position without collision.

**Algorithm 2:** Single-layer fixed-area drone's zone transfer strategy

**Goal:** To transfer or exchange the drones from one zone to another zone without any collision with a fixed transfer/exchange area.

**Premises:** Same as algorithm 1. Additionally, $T_o{}^{LR}$ and $T_o{}^{RL}$ represents $o^{th}$ zone transfer area from left to right zone and from right to left zone respectively.

Allocate one $D_j^i$ in each $Z_i$
**For each** COVID-19 drone-based operation:
    **If** ($D_j^i$ in $Z_i$ and $D_j^{i+1}$ in $Z_{i+1}$ want to swap the location) then
        **If** ($T_o{}^{LR}$ and $T_o{}^{RL}$ are empty) **then:**
            **Move** $D_j^i$ in $Z_i$ to $Z_{i+1}$
            **Move** $D_j^{i+1}$ in $Z_{i+1}$ to $Z_i$
        **End If**
        **Else If** ($T_o{}^{LR}$ is not empty) **then:**
            **Move** $D_j^i$ in $Z_i$ to $T_o{}^{RL}$
            **While** ($T_o{}^{LR}$ is not empty):
                **Wait** $D_j^i$ in $T_o{}^{RL}$
            **End While**
            **Move** $D_j^i$ in $T_o{}^{RL}$ to $Z_{i+1}$
            **Move** $D_j^{i+1}$ in $Z_{i+1}$ to $Z_i$
        **End If**
        **Else If** ($T_o{}^{RL}$ is not empty) **then:**
            **Move** $D_j^{i+1}$ in $Z_{i+1}$ to $T_o{}^{RL}$
            **While** ($T_o{}^{RL}$ is not empty):
                **Wait** $D_j^{i+1}$ in $T_o{}^{RL}$
            **End While**
            **Move** $D_j^{i+1}$ in $T_o{}^{RL}$ to $Z_i$
            **Move** $D_j^i$ in $Z_i$ to $Z_{i+1}$
        **End If**
    **End If**
    **Else**
    Start thermal scanning, social distancing, COVID-19 hotspot detection, sanitization and data analytics after every $\delta$ intervals.
    **End If**

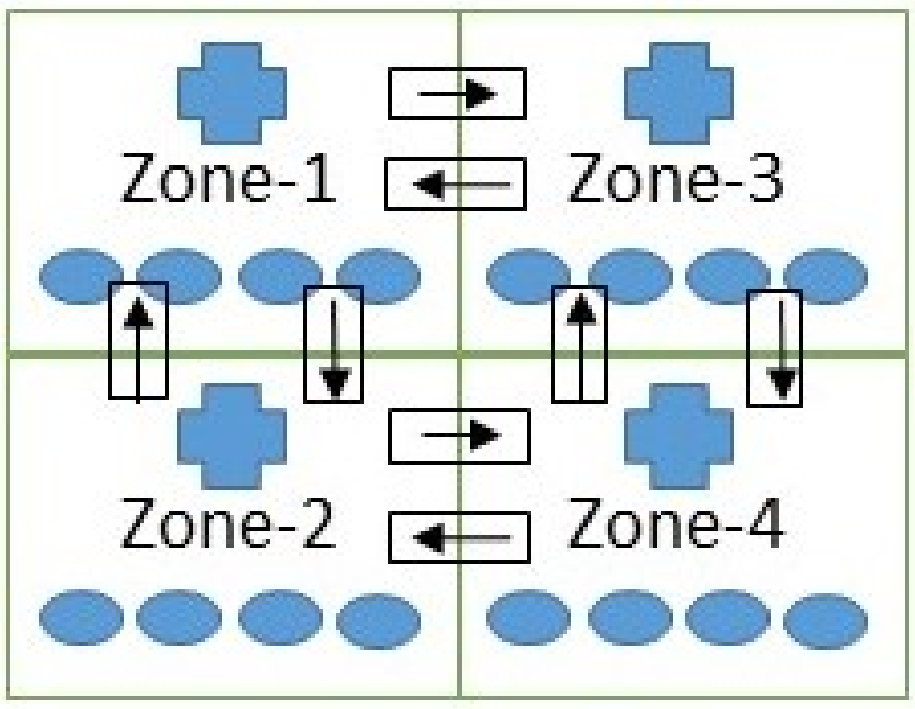

**Figure 5:** Single layer fixed-area zone transfer strategy

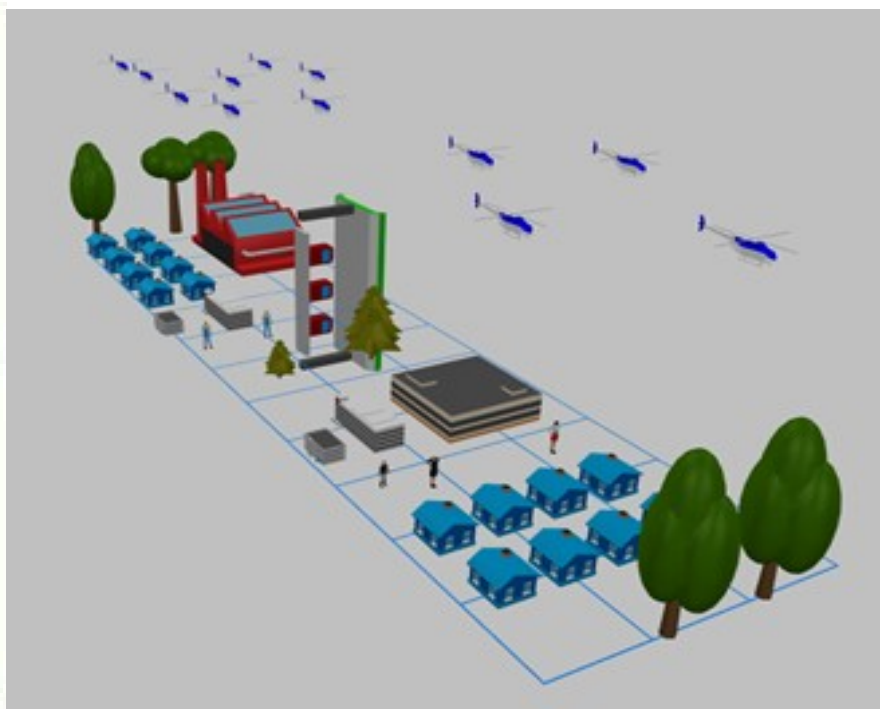

**Figure 6:** Proposed Single-layered Drone-based system for COVID-19 (3D View)

Fig. 7 shows the zigzag movement and zone-transfer strategy. Here, drones enter the area through an entry zone and follow a zigzag strategy for transferring the zone. Algorithm 3 explains this strategy in detail. Additionally, algorithm 4 helps in identifying any drone's current zone by passing the index value. This index value could be mapped to (latitude, longitude) value as well. Further, the collision feasibility zone (as shown in Figure 4) in each zone will avoid small distance collisions. Fig. 8 shows a single-layer drone movement strategy where multiple drones can enter the area through multiple and parallel zone's entry points. Algorithm 5 explains this strategy in detail.

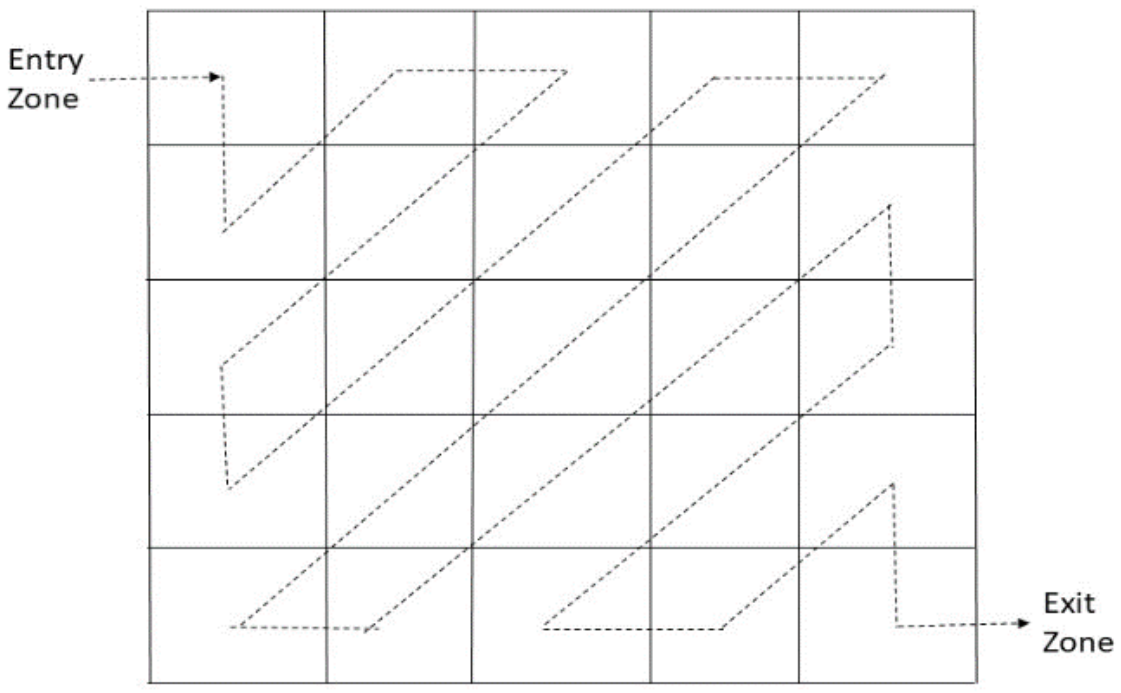

**Figure 7:** Single-layer zig-zag strategy

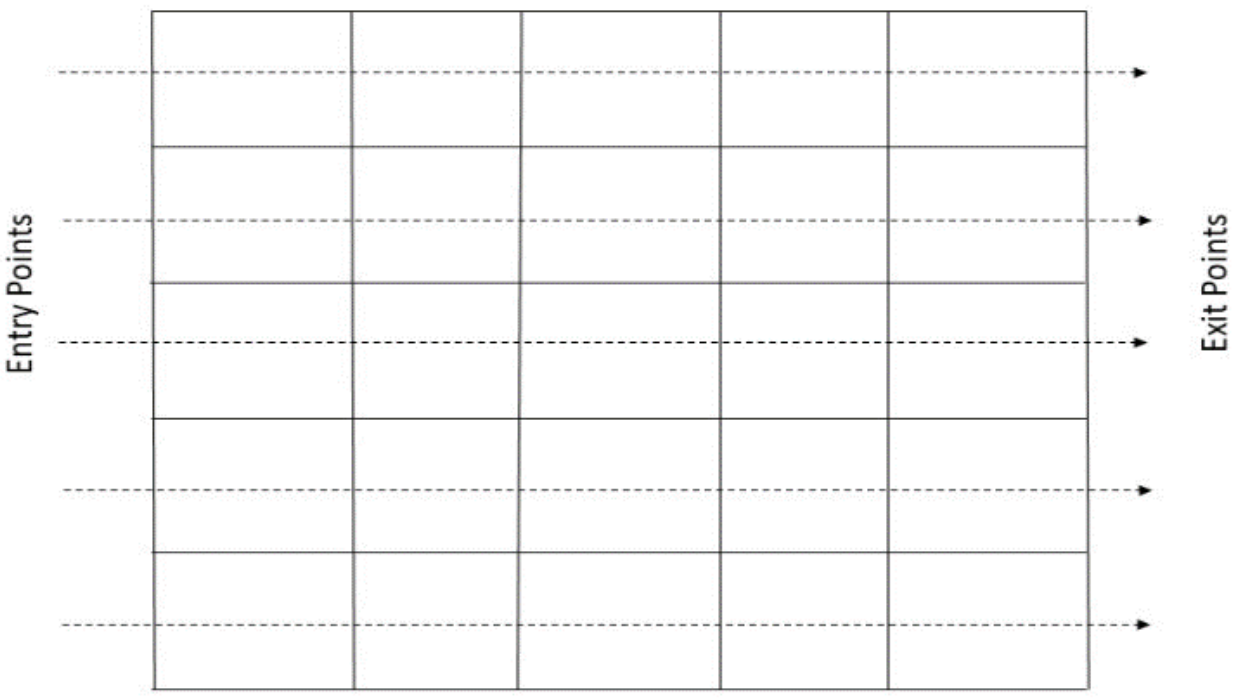

**Figure 8:** Single-layer parallel movements strategy

**Algorithm 3:** Zigzag zone transfer strategy

**Goal:** To transfer or exchange the drones from one zone to another zone in a zig-zag strategy

**Premises:** Same as algorithm 2. Additionally, $(a, b)$ represents the index value in $n \times n$ zone matrix and $\delta$ is an interval between two drone's movements.

**Assumptions:** The given area is divided into $n \times n$-zone matrix and each zone has an area $\tau^2$.

1. Set Index = [ ]
2. **For** $k = 0$ to $n^2 - 1$:
3.     **If** $k \geq n * (n + 1)/2$:
4.         $(a, b)$ = Upper_Zone_Matrix_Index($k$, $n$)
5.     **Else**
6.         $(a, b)$ = Lower_Zone_Matrix_Index($k$, $n$)
7.     **End If**
8.     Append $(a, b)$ to Index
9. **End For**
10. **Set** $j = 0$
11. **For each** $\delta$ interval:
12.     **For** $(a, b)$ in Index:
13.         **If** $(a, b)$ is empty:
14.             Move $D_j^i$ to $(a, b)$
15.         **End If**
16.     **End For**
17.     $j = j + 1$
18. **End For each**

**Function** Upper_Zone_Matrix_Index($k$, $n$)

1. $a, b$ = Upper_Zone_Matrix_Index($n$*$n$-1- $k$, $n$)
2. **return** ($n$-1-$a$, $n$-1-$b$)

**Function Lower_Zone_Matrix_Index(k,n)**

1. $a = (\sqrt{1 + 8 * k} - 1/2$
2. $b = k - a * (a+1) / 2$
3. **If** $a \neq 0$:
4.     **return** $(b, a\text{-}b)$
5. **Else**
6.     **return** $(a\text{-}b, b)$
7. **End If**

**Algorithm 4:** Identify drone's current location in a zig-zag zone transfer strategy

**Goal:** To identify the drone's current zone for operational control

**Premises:** Same as algorithm 3.

**For each** $D_j^i$
    $Z_i$= Drone_Zone_Value($a$, $b$, $n$)
**End For each**

**Function** Drone_Zone_Value($a$, $b$, $n$)

**If** $a + b \geq n$ :
    **return** $n*(n-1)$-$Drone_Zone_Value(n$-1-$i$, $n$-1-$j$, $n$)
**Else**
    $k=(a+b)*(a+b+1)/2$
    **If** $(a+b) \neq 0$:
        **return** $k+a$
    **Else**
        **return** $k+b$
    **End If**
**End If**

**Algorithm 5:** Single layer parallel movement strategy

**Goal:** To transfer or exchange the drones from one zone to another zone in a parallel drone movement strategy.

**Set** $l = 0$
**Set** $a=0$:
**While** (True):
    **For** $b=0$ *to* $n$:
        Move $D_{j+l}^i$ to ($a$, $b$)
        Move $D_{j+l+1}^i$ to ($a+1$, $b$)
        Move $D_{j+l+2}^i$ to ($a+2$, $b$)
        ....
        Move $D_{j+l+n}^i$ to ($a+n$, $b$)
    **End For**
    **If** $\delta$ interval spent:
        $l = l+1$
    **End If**
**End While**

## *4.2 Multi-Layer Zone-Transfer Algorithms*

This section presents two multi-layered zone-transfer algorithms: two-layer zone transfer and hybrid zone-transfer. Fig. 9 shows a two-layered drone transfer strategy. Here, two-layers are proposed to have different drone-based COVID-19 activities (at different layers). The top-layer is considered as a zone-transfer layer and the bottom layer is considered as the COVID-19 operation layer. Now, transfer from one layer to another layer for collision avoidance happens through movements from the operation layer (layer-2) to the zone-transfer layer (layer-1). For example, drones present in the zone-2 and zone-4 want to swap their positions. Thus, zone-2 and zone 4 drones will move from layer-2 to layer-1. Thereafter, zone-3 drone from layer-1 will move to zone-2 of layer-2, and zone-1 drone from layer-1 will move to zone-4 of layer-2. This approach can be extended to multiple layers for different COVID-19 or other services' operations/activities and/or working on uneven building structures. Every second layer in the multi-layered architecture is considered as zone-transfer layer. Algorithm 6 explains the multi-layer drone transfer strategy in detail. Fig. 10 and Fig. 11 show the simulation of the two-layered drone-based approach in 2D and 3D views respectively (in execution).

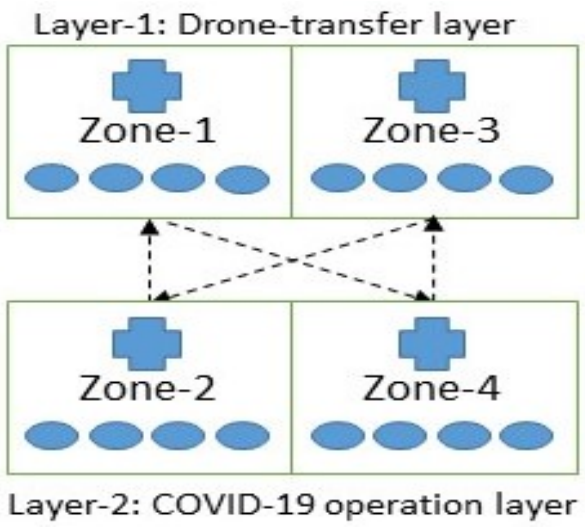

**Figure 9**: Two-layered zone transfer strategy

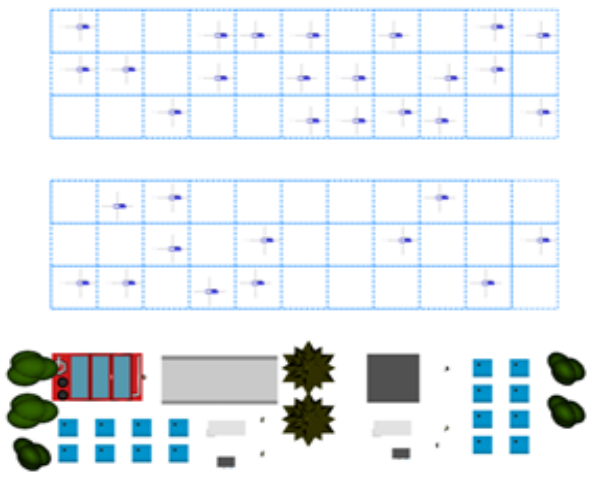

**Figure 10:** Proposed Two-layered drone-based system for COVID-19 (2D View)

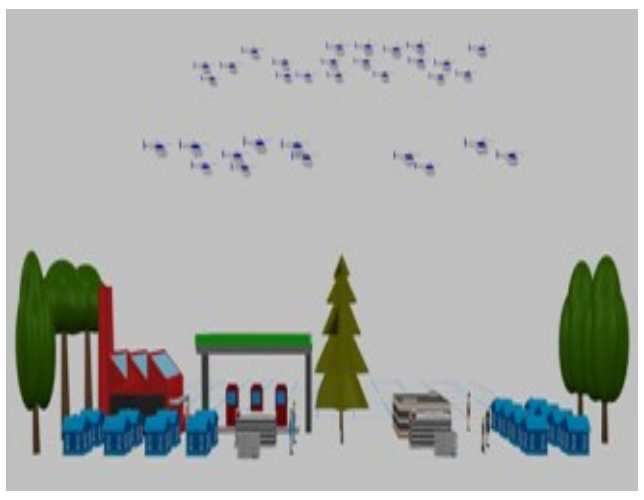

**Figure 11:** Proposed Multi-layered Drone-based system for COVID-19 (3D View)

**Algorithm 6:** Multi-layered zone transfer strategy

**Goal:** To transfer or exchange the drones from one zone to another zone without any collision with the usage of multi-layered air zones

**Premises:** Same s algorithm 1. Additionally, $L_m$ represents $m^{\text{th}}$ layer.

1. **For each** COVID-19 drone-based operation:
2.     **If** ($D_j^i$ in $Z_i$ and $D_j^{i+1}$ in $Z_{i+1}$ at $m^{\text{th}}$ layer want to swap the location) then
3.         **Move** $D_j^i$ in $Z_i$ at $m^{\text{th}}$ layer to $Z_i$ at $(m\text{-}1)^{\text{th}}$ layer
4.         **Move** $D_j^{i+1}$ in $Z_{i+1}$ at $m^{\text{th}}$ layer to $Z_{i+1}$ at $(m\text{-}1)^{\text{th}}$ layer
5.         **Move** $D_j^i$ in $Z_i$ at $(m\text{-}1)^{\text{th}}$ layer to $Z_{i+1}$ at $m^{\text{th}}$ layer
6.         **Move** $D_j^{i+1}$in $Z_{i+1}$ at $(m\text{-}1)^{\text{th}}$ layer to $Z_i$ at $m^{\text{th}}$ layer
7.     **End If**
8.     **Else If (**$D_j^i$ in $Z_i$ at $m^{\text{th}}$ layer want to move to an empty drone zone $Z_{i+1}$) then
9.         **Move** $D_j^i$ in $Z_i$ at $m^{\text{th}}$ layer to $Z_i$ at $(m\text{-}1)^{\text{th}}$ layer
10.         **If** (no other neighboring zone's drone want to moveto this zone) **then**
11.             **Move** $D_j^i$ in $Z_i$ at $(m\text{-}1)^{\text{th}}$ layer to $Z_{i+1}$ at $m^{\text{th}}$ layer
12.         **Else**
13.             **Move** first request drone to $Z_{i+1}$ at $m^{\text{th}}$ layer
14.         **End If**
15.     **If** ($D_j^i$ in $Z_i$ and $D_j^n$ in $Z_n$ at $m^{\text{th}}$ layer want to swap the location) then
16.         **Move** $D_j^i$ in $Z_i$ at $m^{\text{th}}$ layer to $Z_i$ at $(m\text{-}1)^{\text{th}}$ layer
17.         **Move** $D_j^n$ in $Z_n$ at $m^{\text{th}}$ layer to $Z_i$ at $(m\text{-}1)^{\text{th}}$ layer
18.         Apply single layer drone's zone transfer strategy proposed in algorithm 2 and find the route from $Z_i$ to $Z_n$ for transfer
19.     **End If**
20. **End for**

Fig. 12 shows the multi-layer hybrid movement's strategy. In this strategy, the drones' movement area is divided into multiple layers as per the existing infrastructure and possibilities of divisions. Thereafter, different zone movement and transfer strategies could be applied for different operations. For example, four levels with four areas are shown in Fig. 12. Area-2 at level-2 and area-4 at level-4 can use two-layered drone-movement and zone transfer strategy for COVID-19 operations. Whereas, area 3 at level-3 can use a parallel drone movement strategy for passing by drones. Similarly, area-1 at level-1 can use zig-zag drones' movement strategy for product delivery.

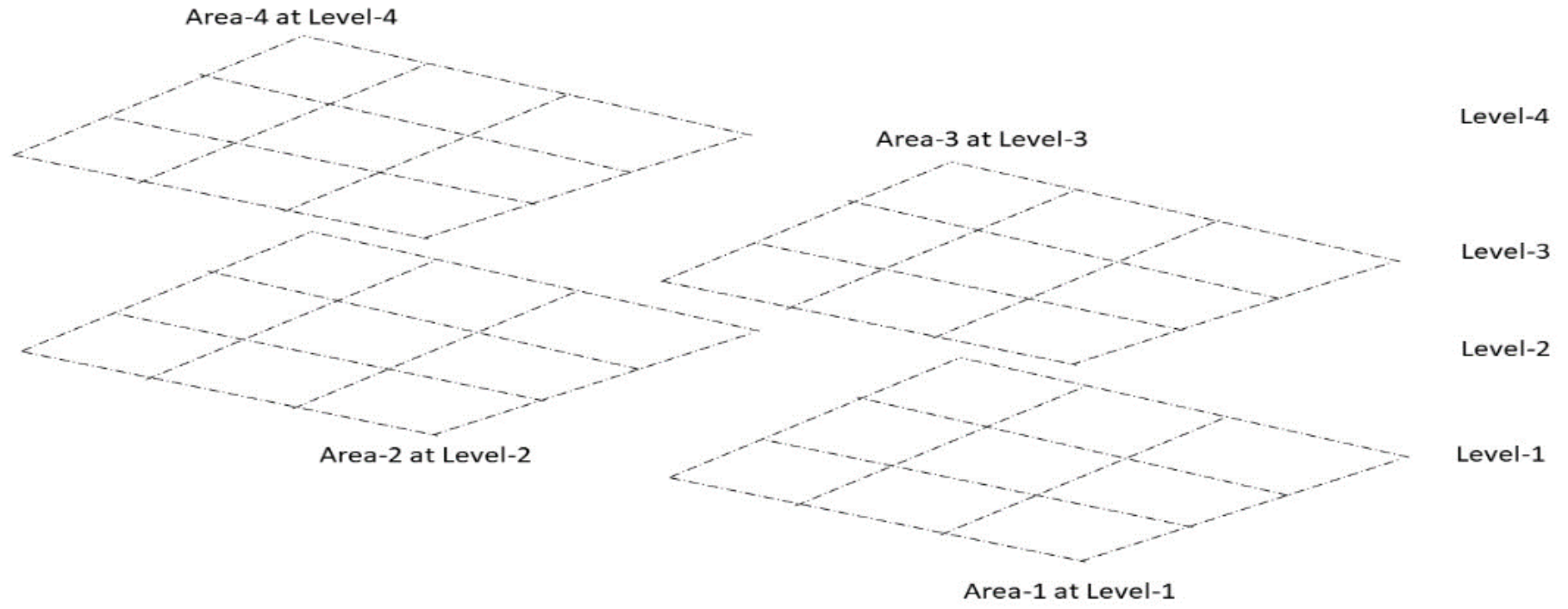

**Figure 12:** Multi-Layer Hybrid Movements Strategies

### *4.3 Social-Distancing Algorithms*

This section shows two different approaches to maintain social distancing in COVID-19 operation. This system helps in generating alerts/alarms through drones when social-distance is not maintained in an observational area. Algorithm 7 proposed an approach that ensures the symmetric distance between any two persons standing in a queue. Algorithm 8 proposed an approach that ensures the symmetric distance between any two persons when the population is randomly distributed in a geographical region. Algorithm 7 and Algorithm 8 uses the following four functions of COVID-19 social distancing.

- *Max_Person_Calculation():* This function measures the number of people allowed in a geographical area. If the number of people is greater than a certain threshold then it intimates the people through a wearable sensor or speaker based alert generation system for spreading out.
- *Distance_Measurement():* This function calculates the distance between two persons using various distance measurement formulas. Initially, the distance is recommended to be measured using acoustic, acoustic, laser, radio, infrared or mono/stereo distance sensors but if none of these choices are available then it measures the distance from longitude and latitude measurements. For example, tunnel-based distance measurements are preferred only for those cases when the distance between two persons is very long. In COVID-19 cases, tunnel-based distance formula can be used for data analytics in two different places and then comparing the results.
- *Person_Intimation():* This function uses the intimation system to people through wearable sensors or alert generation based system.
- *Control_Room_Notification():* this function is used to measure the functionalities of an individual drone as well as the drone-based network. Further, this function measures the COVID-19 social distancing and hotspot identification as well.

**Algorithm 7:** Symmetric distance between any two persons standing in a queue.

**Goal:** To measure the distance between two persons standing consecutively in a queue, ensure minimum distance and intimate through a wearable device, if necessary.

**Premises:** Let $P_i^j$ represents $i^{th}$ person standing in $j^{th}$ queue. Where $i \in \{1,2,\dots,n\}$ and $j \in \{1,2,\dots,m\}$ i.e. there is a maximum of $n$-persons and $m$-queues. $D_{i,k}^j$ is the distance between $i^{th}$ person and $k^{th}$ person when standing in $j^{th}$ queue. Here, $k=i+1$ or $i-1$. Let $\emptyset_{P_i^j}$ and $\emptyset_{P_k^j}$ represents the latitude, and $\lambda_{P_i^j}$ and $\lambda_{P_k^j}$ represents the longitude of $i^{th}$ person and $k^{th}$ person respectively. $R$ represents the radius of the earth, $\boldsymbol{e}(D_{i,k}^j)$ is the maximum error in computing $D_{i,k}^j$. $F$ is the footpoint width of the camera image taken over the earth, $d$ is the ground sample distance and it is used to measure the distance between two adjacent pixels values, $\vartheta$ and $\mu_w$ represents the pixel width in image and image width respectively, $P_i^j(r)$ is the distance between the camera and person/object, $P_i^j(L)$ represents the length of $i^{th}$ person standing in $j^{th}$ queue. $C_i^j$ represents $i^{th}$ control room for $j^{th}$ queue. $\delta$ is drone utilization measurement.

```
For j=1 to m:
    Max_Person_Calculation()
    Distance_Mesurement()
```

4.     Person_Intimation()
5.     Control_Room_Notification()
6. **End For**

**Max_person_calculation()**

1. Set count=0
2. **If** (acoustic, laser, radio, infra-red, mono/stereo distance sensors are available) **then**
3.     Measure $P_i^j(r)$
4.     **For each** $P_i^j(r)$:
5.         *Count=count+1*
6.     **End For**
7. **End If**
8. **Else**
9. **For** each object in camera image**:**
10.     **Measure** $P_i^j(L) = \emptyset * P_i^j(r)$
11.     **If** $P_i^j(L) >$ threshold **then**
12.         *Count=count+1*
13.     **End If**
14. **End For**

**Distance_Measurement( )**

1. **If** (distance measurement is based on latitude-longitude and tunnel formula) **then**
2.     **If** (distance_formula is based on spherical_surface having very long distance) **then**
3.     $\Delta X = \cos(\emptyset_{P_k^j}).\cos(\lambda_{P_k^j})$- $\cos(\emptyset_{P_i^j}).\cos(\lambda_{P_i^j})$
4.     $\Delta Y = \cos(\emptyset_{P_k^j}).\sin(\lambda_{P_k^j})$- $\cos(\emptyset_{P_i^j}).\sin(\lambda_{P_i^j})$
5.     $\Delta Z = \sin(\emptyset_{P_k^j})$- $\sin(\emptyset_{P_i^j})$
6.         Tunnel Distance $(T_D) = \sqrt{(\Delta X)^2 + (\Delta Y)^2 + (\Delta Z)^2}$
7.         $D_{i,k}^j = T_D . R$
8.         **If** $(D_{i,k}^j \ll R)$ **then**
9.             $\boldsymbol{e}(D_{i,k}^j) = D_{i,k}^j (D_{i,k}^j / R)^2 / 24$
10.     **End If**
11. **End If**
12. **If** (distance measurement is based on latitude-longitude only) **then**
13.     $D_{i,k}^j = 111.32 * \sqrt{(\emptyset_{P_k^j} - \emptyset_{P_i^j})^2 + (\lambda_{P_k^j} - \lambda_{P_i^j})^2}$
14. **End If**
15. **If** (distance measurement is based on image processing and ground sample distance (GSD)) **then**
*16.*     *d=(g*a*100)/(f* $\vartheta$)*
*17.*     *w=d**$\mu_w$
*18.*     Apply linear discriminant analysis for feature-based person detection in an image with width *w and compute* $D_{i,k}^j$.
*19.* **End If**
20. **If** (distance measurement is based on image processing) **then**
21.     $P_i^j(r) = P_i^j(L)/\emptyset$
22.     $D_{i,k}^j = P_i^j(r) - P_k^j(r)$
23. **End If**

**Person_Intimation( )**

1. **For each** $D_{i,k}^j$:
2.     **If** $D_{i,k}^j <$ threshold **then**
3.         Send signal to $P_i^j(r)$ and $P_k^j(r)$
4.         **Call** Distance_Measurement( )
5.     **End If**

6. **End For**

**Control_Room_Notification( )**

1. **For each** drone:
2. **Measure** $\delta$
3. **If** $\delta \geq$ upper_threshold **then**
4. **Call** drone back
5. **Else If** $\delta <$ lower_threshold **then**
6. **Instruct** to start COVID-19 process
7. **End If**
8. **End For**
9. **For each** $C_i^j$:
10. **For** each $j$ in $C_i^j$:
11. **If** $D_{i,k}^j <$ threshold **then**
12. Send signal to $P_i^j(r)$ and $P_k^j(r)$
13. **Call** Distance_Measurement( )
14. **End If**
15. **End For**
16. **End For**

**Algorithm 8:** Symmetric distance between any two persons when the population is randomly distributed in a geographical region.

**Goal:** To measure the distance between two persons standing consecutively when people are randomly distributed over a geographical region and intimate through wearable devices, if necessary.

**Premises:** same as algorithm 1

1. **For** j=1 to m:
2. Max_person_Calculation()
3. Distance_Measurement() //Same as in algorithm 1
4. Person_Intimation() //Same as in algorithm 1
5. Control_Room_Notification() //Same as in algorithm 1
6. **End For**

**Max_person_Calculation()**

1. Set *count*=0
2. **For each** direction (north, south, west, east) in a circle:
3. **Rotate** sensors in all directions
4. **If** (acoustic, laser, radio, infra-red, mono/stereo distance sensors are available and detect the person) **then**
5. *count=count+1*
7. **Else**
8. **Call** Distance_Measurement() and Measure $P_i^j(r)$
9. **End If**
10. **End For**
11. **For each** $P_i^j(r)$:
6. *Count=count+1*
7. **End For**
8. **For** each object in camera image**:**
9. **Measure** $P_i^j(L) = \emptyset * P_i^j(r)$
10. **If** $P_i^j(L) >$ threshold **then**
11. *Count=count+1*
12. **End If**
13. **End For**

## 5. Performance Evaluation

This section discusses real-time and simulation-based realization of proposed system architecture incorporating algorithms. It presents these evaluation cases adopted for COVID-19 scanning, sanitization, monitoring, analysis, and statistics for the control room.

### *5.1 Case 1: Drone-based Real-Time System for COVID-19*

We present a real-time drone system for COVID-19 developed by the Indian Robotics Solution (IRS)**.** It is used for sanitizing, monitoring, and controlling COVID-19 pandemic in Delhi and the National Capital Region (NCR) of India.

• ***Thermal Corona Combat Drone (TCCD)*****:** This system is deployed in-practice for object identification and thermal scanning over many areas in Delhi and NCR, and found it as a suitable help to fight against COVID-19. This is real-time implementation. Table 3 shows the complete specifications and features of the build and used a TCCD system [6].

**Table 3.** Thermal Corona Combat Drone (TCCD)

| Parameters | Value |
|---|---|
| Type of drone | Multi-Rotor Drone |
| No. of Rotor | Six rotors (Hexa-copter) |
| Frame Material | Carbon fiber Sheet and tube |
| Camera | Thermal camera and RGB camera with Spot Light for Night Operation |
| Remote Control Working frequency | 2.4Ghz and 5.8Ghz |
| Circuit designing | In-house |
| Flight Mode | Autonomous mode suing waypoint, Manual Mode, Hovering |
| Coding skills and Technology | Thermal imaging, Geothermal sensing, GPS tracking, Compaction and Photography |
| Model | Thermal Corona Combat Drone (TCCD) |
| Machinery Used | Waterjet, CNC, 3D printer, Metal moulding, and R & D tools |
| Spraying System | Brushless Pump + 4 nozzle |
| Gimbal | 3 Axis Gimbal for camera movement |
| Battery Type | Lithium Polymer batteries (1 battery set) |
| Battery cycle | 200 cycle/each |
| Charger | Lipo Balance charger (20A) |
| Payload | 10 kg payload |
| Attachment | Medicine box 2 kg, 5L sanitizing tank, 2 cameras, Loud Speaker |
| Weight of drone | 15 Kg. (Empty tank) and 20 Kg. (full tank) |
| Satellites for GPS | A cluster of 60 satellites |
| Flight time | 35- 40 minutes (if only thermal imaging)<br>12-15 minutes (if spray and thermal imaging) |
| Success rate | 95% |

• ***Working of TCCD*****:** TCCD takes off vertically from the ground and no explicit launching pad is required. First, the team finalizes the area to be surveyed (densely populated areas usually) and takes clearances from relevant government authorities. When the drone is at a suitable height inside a society (suburban area), it makes announcements requesting the residents to come out on their balconies. Once the people are standing in their balconies then thermal scanning begins for one person at a time where the body temperature is recorded. Thermal scanning can be done of a person within 20 feet range. If the recorded temperature of any person is greater by at least 2$^{o}$C than normal body temperature, an alarm beeps and an announcement is made over the loudspeaker fitted on the drone. To take an action the person having higher temperature may come down otherwise an official from the outreach team may go up for more check-ups. A high temperature is one of the common symptoms of coronavirus and temperature screening is used at many international borders and hospitals for initial symptomatic detection [3]. This temperature measurement is followed by the healthcare system where other COVID-19 symptoms including nausea, headaches, fatigue, loss of taste or smell etc. are identified. Further, immunodiagnostic tests are performed for confirming COVID-19. This multipurpose UAV facilitates society by sanitizing the entire area via built-in spray functionality and providing medical supplies (if required). Two cameras help in thermal image sensing and a video recording of the survey site even in the night using night vision mode. Thermal image sensor recording and video of live scanning are visible on a connected portable digital assistant like a mobile phone or phablet.

• **Outcomes:** Testing of TCCD has been completed in more than 15 densely populated areas (mostly slum areas) where the roads are not good and reachability is an issue for the officials. Around 1 million people live in these areas (Delhi, India), which include Keshav Puram, Narela, Sadar Bazaar, Slum areas of Civil Lines, Majnu ka Tila, Paschim Vihar, and Dwarka. A sample of the scanned images, which can be seen on the PDA device, is shown in Fig. 13. Fig. 13(a) and Fig. 13(b) show the temperature reading measurement over a PDA device from a long and short distance respectively. Fig. 13(c) shows the drone designed and used for COVID-19 operations. Fig. 13(d) shows the thermal image-based surface scanning. Scanning and sanitization of these areas has been done in just 7 days. It is observed that an area of 2 km radius can be sanitized within 10 minutes. Thereafter, IRS teams scaled-up this experimentation to a large area with proposed artificial intelligence and fast image processing. Further, the experimentation is enhanced by using electrostatic spray. The electrostatic spray helps droplets of sanitizer to be more centric and overlapping thus making it more effective and avoids wastage.

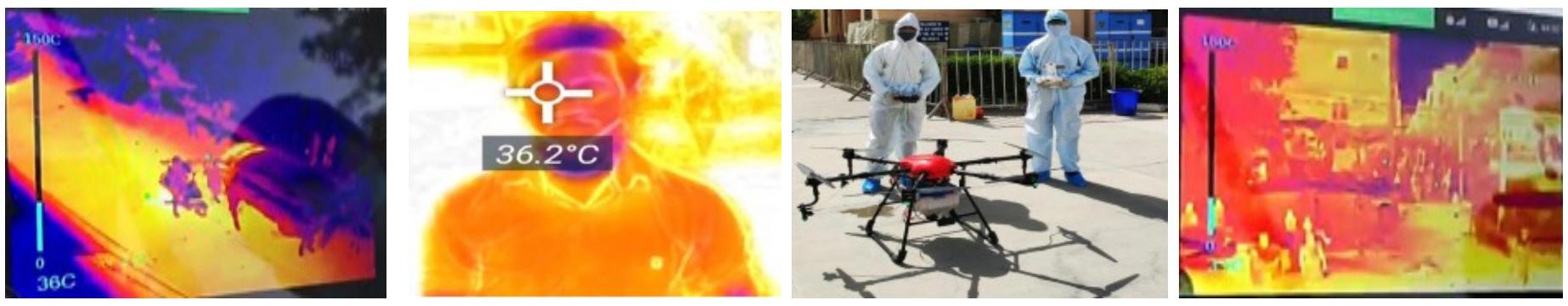

**(a):** Thermal Image-based Temp. Measurement (at a long-distance)

**(b):** Thermal Image-based Temp. Measurement (at a close distance)

**(c):** Drone designed

**(d):** Thermal Image-basec surface and area scanning

**Figure 13:** Real-Time Drone-based COVID-19 Temperature Reading, Thermal Scanning and Medication System

## 5.2 *Case 2: Drone-based Simulation for COVID-19 Operations (Monitoring, Sanitising, and Thermal Imaging ) in Outdoor Slum or Overused Areas*

We have developed a simulation-based drone system for COVID-19 using AnyLogic [31] and JaamSim Simulators [32]**.** In this simulation, multimedia modeling, agent-based modeling, discrete event modeling, and system dynamics are applied over an area variation of 250m$^2$ to 1000m$^2$. Fig. 14 shows the drone-based simulation when people are living or moving in close locations like in slum areas. The chances of sharing toilets, bathrooms, water supply, and other public resources are very high. Thus, this increases the chances of COVID-19 pandemic as well. To avoid spreading of COVID-19 cases, regular monitoring and sanitization are required. Fig. 14 shows the side view of the area where drones are used for sanitizing the space. Here, two drones are shown that can move freely and sanitize the area with instructions. Fig. 15 shows the drone-based thermal image of people's movement monitoring and image processing. Here, people's movements are monitored and density-based analysis is performed with the help of proposed multi-layered architecture. The high-density movement area needs frequent sanitization as compared to low-density movement area. Fig. 15 shows the single-day observations and severity of experimentation increases with the darkness of the red color. This indicates that sanitization is required at those places. Fig. 16 shows the circuit used for simulating drone-based COVID-19 operations. Various components of this circuit are briefly explained as follows.

- pedSource: ensures (i) significant passengers travel, and (ii) passengers increase with services and type of services.
- pedGoTo: ensures randomness and free-pedestrian movements.
- pedWait: ensures (i) significant and random stay-time, and (ii) social-distancing.
- pedSink: ensures smooth pedestrian removal.

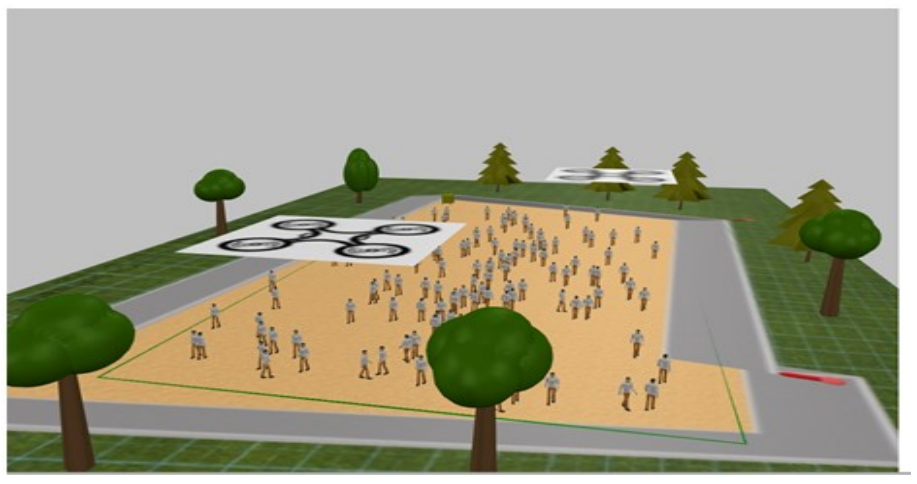

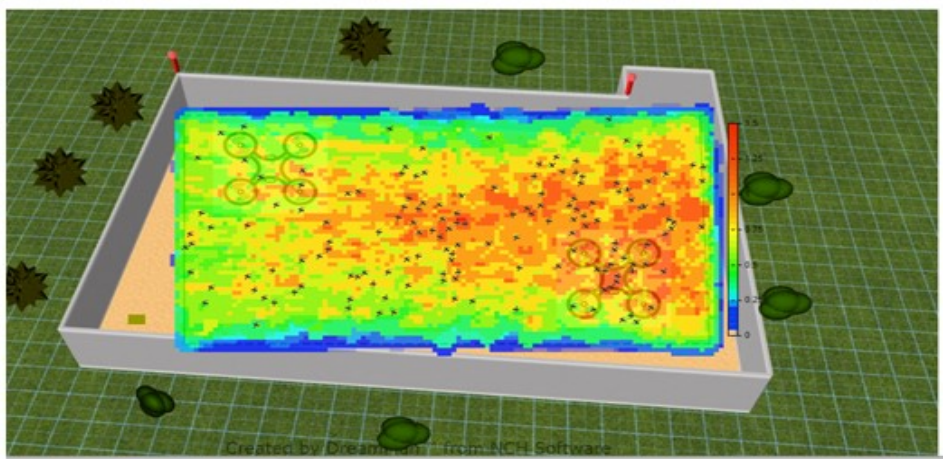

**Figure 14:** Drone-based pandemic areas considered for monitoring (side-view).

**Figure 15:** Drone-based thermal image for people movement monitoring and sanitizing in COVID-19 hotspots. (single-day observations)

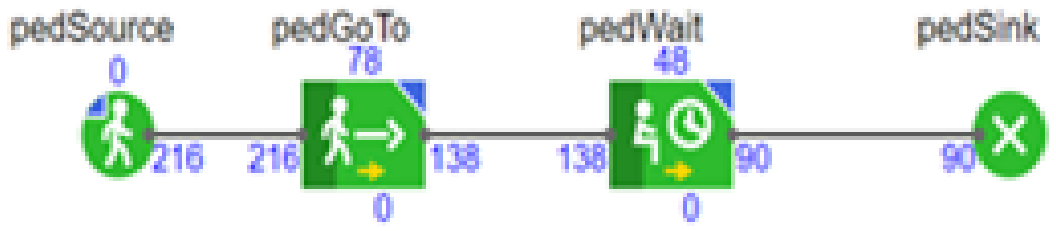

**Figure 16** Anylogic Simulation Model for COVID-19 Operation

Fig. 17 shows the comparative analysis of the time required to sanitize 100 to 1200 kilometers of the area with variations in the number of drones. Results show that 18900, 9390, 3680, and 2293 minutes are required to cover 1200 kilometers of the area with 3, 10, 20, and 30 drones respectively. The drone recharging and sanitizer filling time are additional.

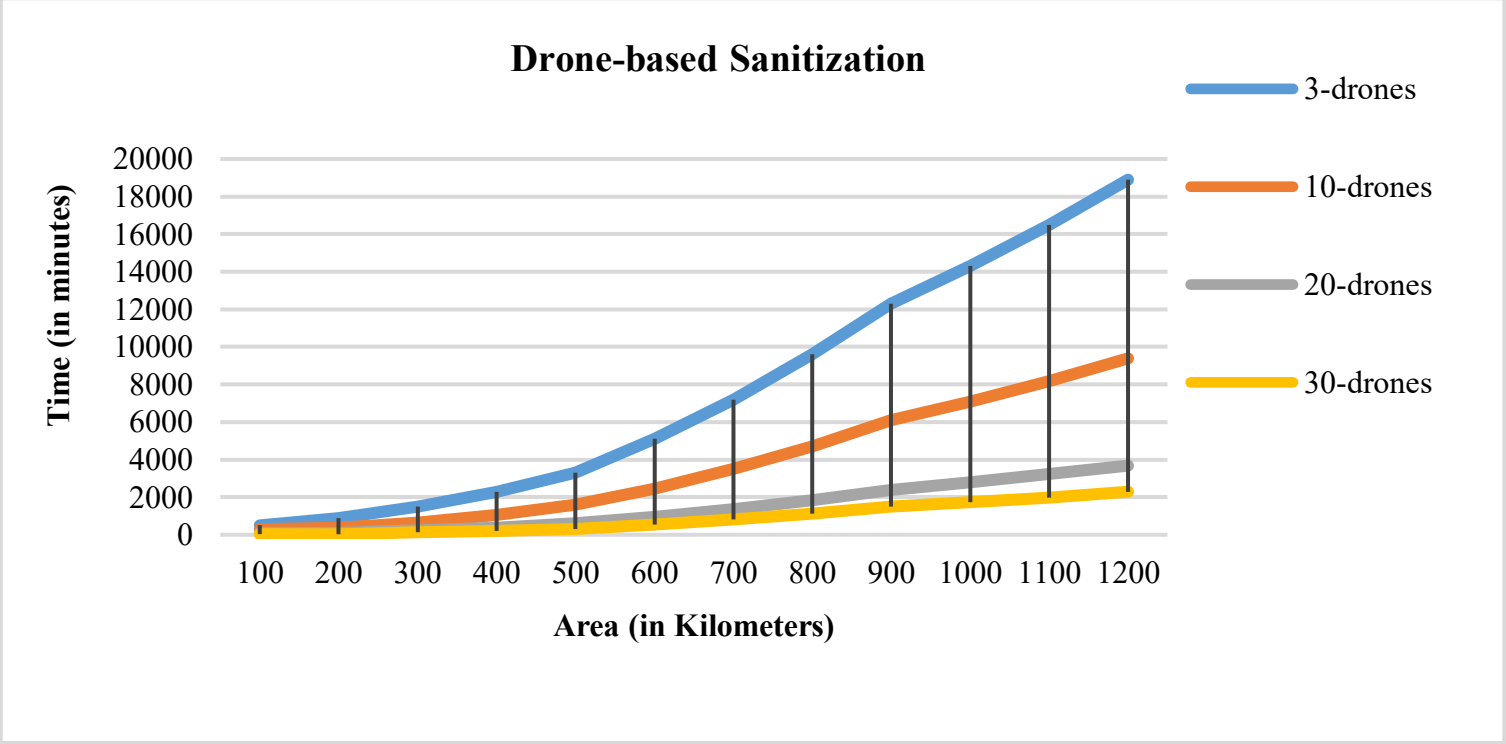

**Figure 17:** Comparative analysis of the time required to sanitize 100 to 1200 kilometers of the area with variations in the number of drones.

## 5.3 *Case 3: Drone-based Simulation for COVID-19 Operations in Indoor Monitoring*

Fig. 18 shows the experimentation for drone-based simulation for inspecting indoor COVID-19 patients. This is a simulation-based study of indoor COVID-19 operations. In an indoor activity, nano or low altitude drones are preferred because of their various advantages [29] [30]. Fig. 18 shows the hospital's internal building and multiple patient wards for admission. This drone's camera is programmed for inspecting the patients based on their movements and density. It is assumed that irrespective of drone-based special sanitization service, the wards are sanitized (manually) after regular intervals or after patient discharge. Thus, drone-based sanitization is made mandatory in those areas where the movement of people/patients is higher.

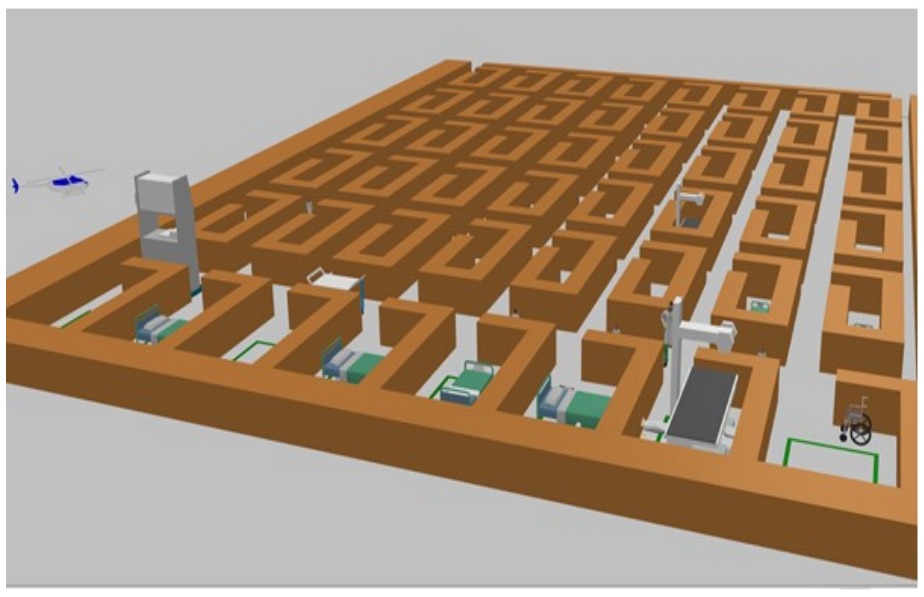

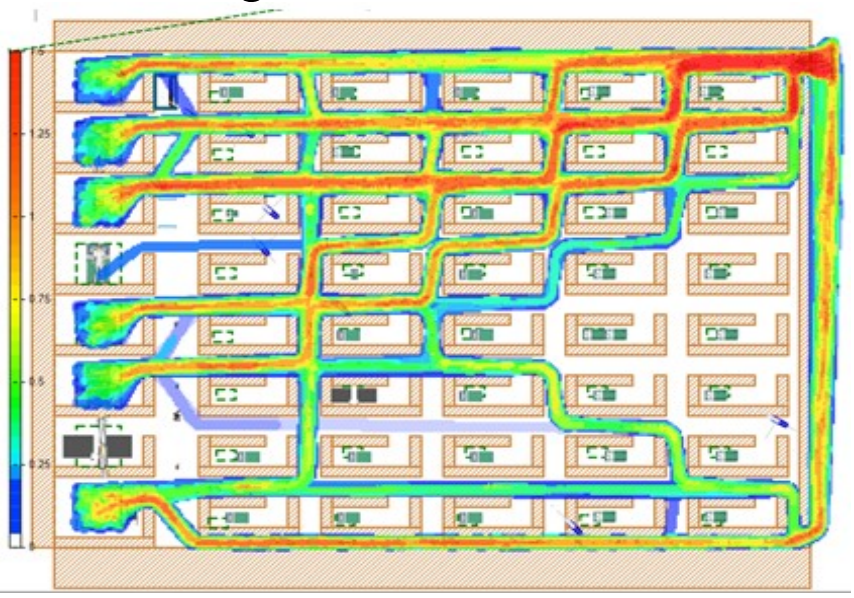

**Figure 18:** Hospital building for inspection (3D view)

**Figure 19:** AnyLogic Model for inspecting COVID-19 patients and sanitization

The drone camera-enabled hospital room gives real-statistics to proposed architecture for data analysis and instructions are given back to the drone for the operation of sanitization and medication in the predefined area.

Drone's camera used for COVID-19 operations shows that the accuracy of the proposed simulation is higher (93% approx.). Fig. 19 shows the simulation's thermal image of density-based areas that need sanitization. An increase in darkness of red color indicates urgent sanitization.

### 5.4 *Case 4: Drone-based Simulation for Social Distancing*

This section presents the drone-based simulation of social distancing experiments. In the experiments, Algorithm 7 is used for ensuring symmetric distance between two persons. Algorithm 7 computes the number of persons standing in a geographical area in its initial calculations. If the number of people standing in an area is more than the acceptable limit then an alert is generated to reduce the count or stop the services. Thereafter, a distance measurement process starts. Now, distance can be measured in multiple ways and through different formulas. Fig. 20 shows the application of the proposed approach in simulation. Fig. 20 (a) shows a road where people are free to move through the front-view camera. In this view, three drones are shown to monitor the movement of people. Fig. 20(b) shows social distancing in practice and the queue formation process. Here, people have made a queue and they are served for their necessity as well. This queue ensures that if certain people are not following the social distance rules then they will not be served and they have to go back to the queue-end for their turns. Fig. 20(c) is the circuit diagram designed and programmed for social distancing simulations. Various components of this diagram are explained as follows.

- *pedSource:* This generates the number of pedestrians for experimentation from a pre-defined and fixed-line. The pedestrians are free to move randomly and in any direction in the pre-defined area. However, everyone has to follow the queue and social distancing otherwise they have to go back at the start of the queue and follows the current position.
- *atFareGates:* As the number of pedestrians increases with time but space does not allow to move above a fixed number of passengers. Thus, a crowd is generated at the entry gates. To avoid the pandemics, all people are alerted to go back or follow the social distancing here as well.
- *pedGoTo:* specifies the number of pedestrians that are moving from the entry gate to service point. This component does not allow multiple people to serve at a time. However, the services resumed only if all are following the social distancing experiment.
- *pedService:* This component ensures that each person following the social distancing experiment is served through a queue. The serving and time duration is fixed and pre-defined.
- *pedSink:* This component removes the pedestrians from observational space after they are served and at the position of the exit gates.

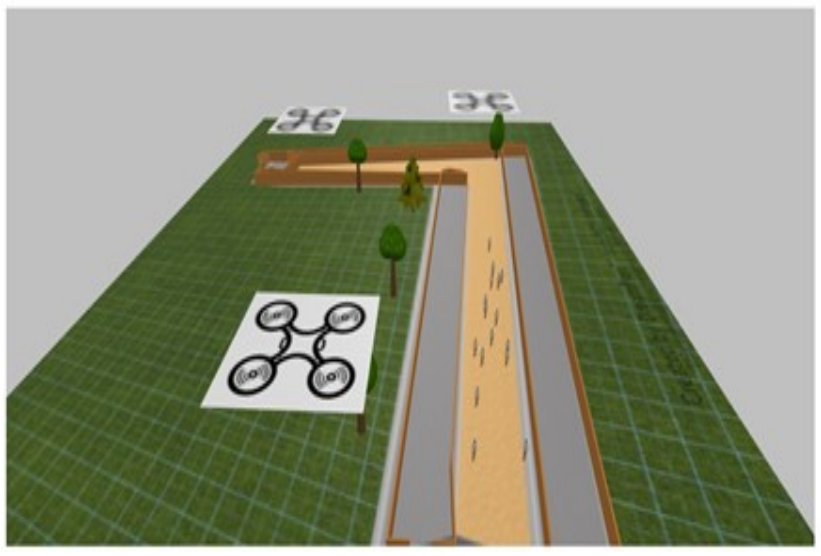

**(a)** Drone-based Monitoring and Simulation (front-view)

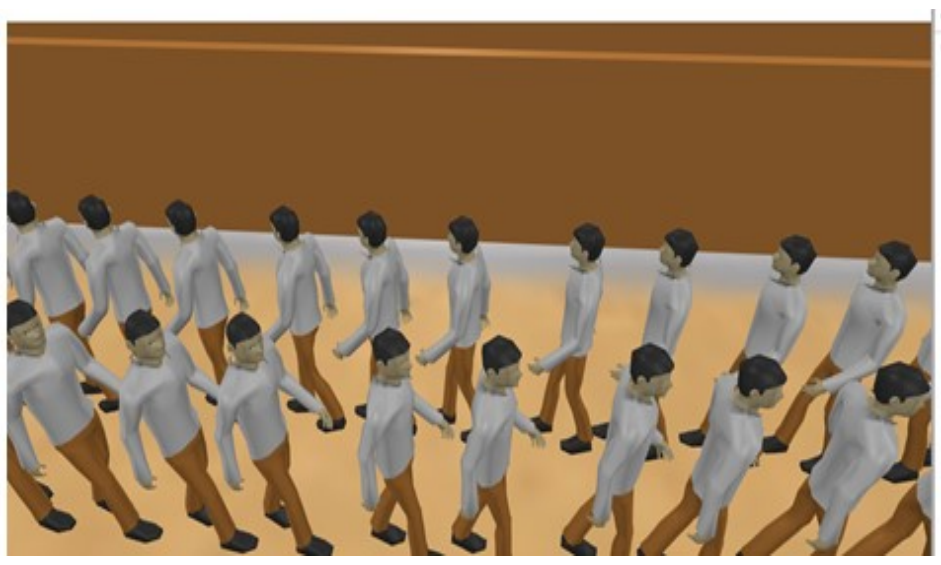

**(b)** Drone-based social distancing queue

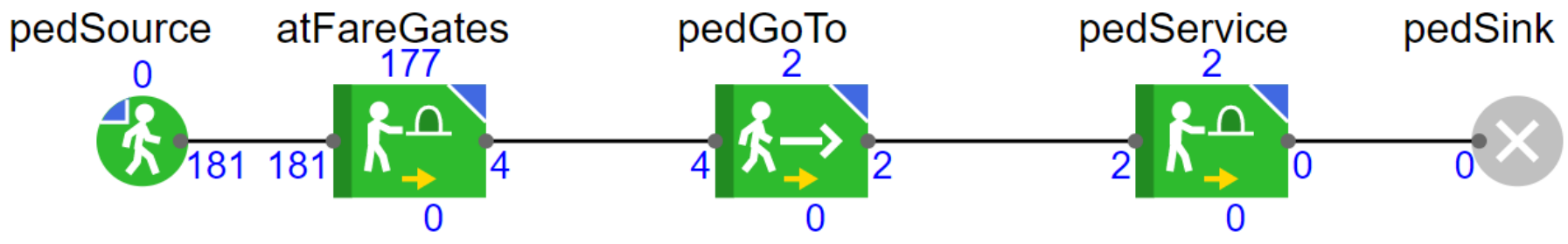

**(c)** Drone-based AnyLogic Model for Social Distancing

**Figure 20.** Drone/UAV-based Simulation Model for Social Distancing in hospitals, banks, supermarket etc.

Fig. 21 shows the comparative analysis of the number of persons checked for social distancing with variations in the number of drones. In this experimentation, 3, 10, 20 and 30 drones are taken for checking the social distancing. Results show that the number of persons checked with drones over simulation time variations increases exponentially. Results show that around 3389 persons can be checked in 55 minutes with 3 drones. Similarly, 13398, 16298, and 19697 persons can be checked in 55 minutes with 10, 20 and 30 drones respectively. Further, Fig. 22 shows a comparative analysis of the number of persons served with medicine supply after a social distancing check. It is observed that 1612, 10073, 13129, and 16166 persons can be served along with social distance checking when the maximum number of working parallel medicine supply units=20, and the maximum time taken to supply a medicine= 120 seconds.

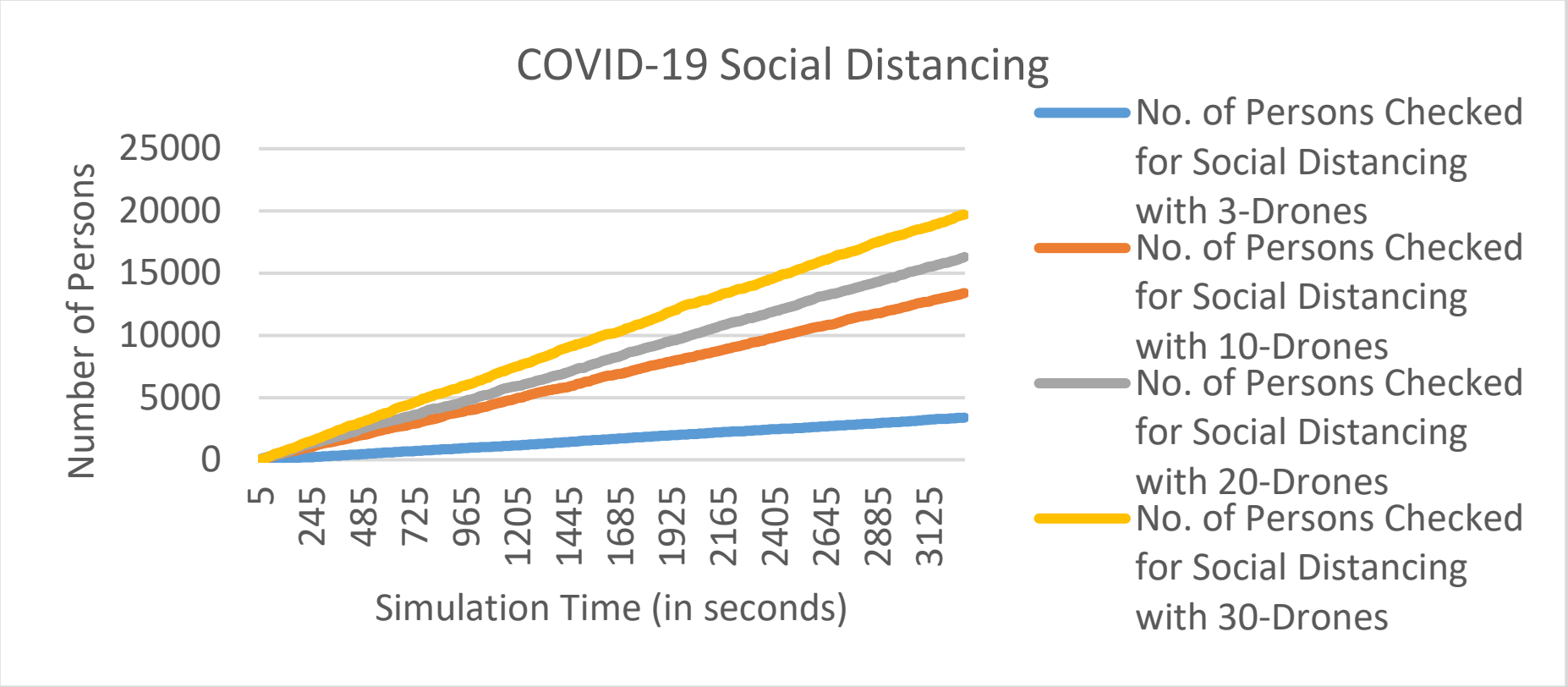

**Figure 21:** Comparative analysis of the number of persons checked for social distancing with variations in the number of drones

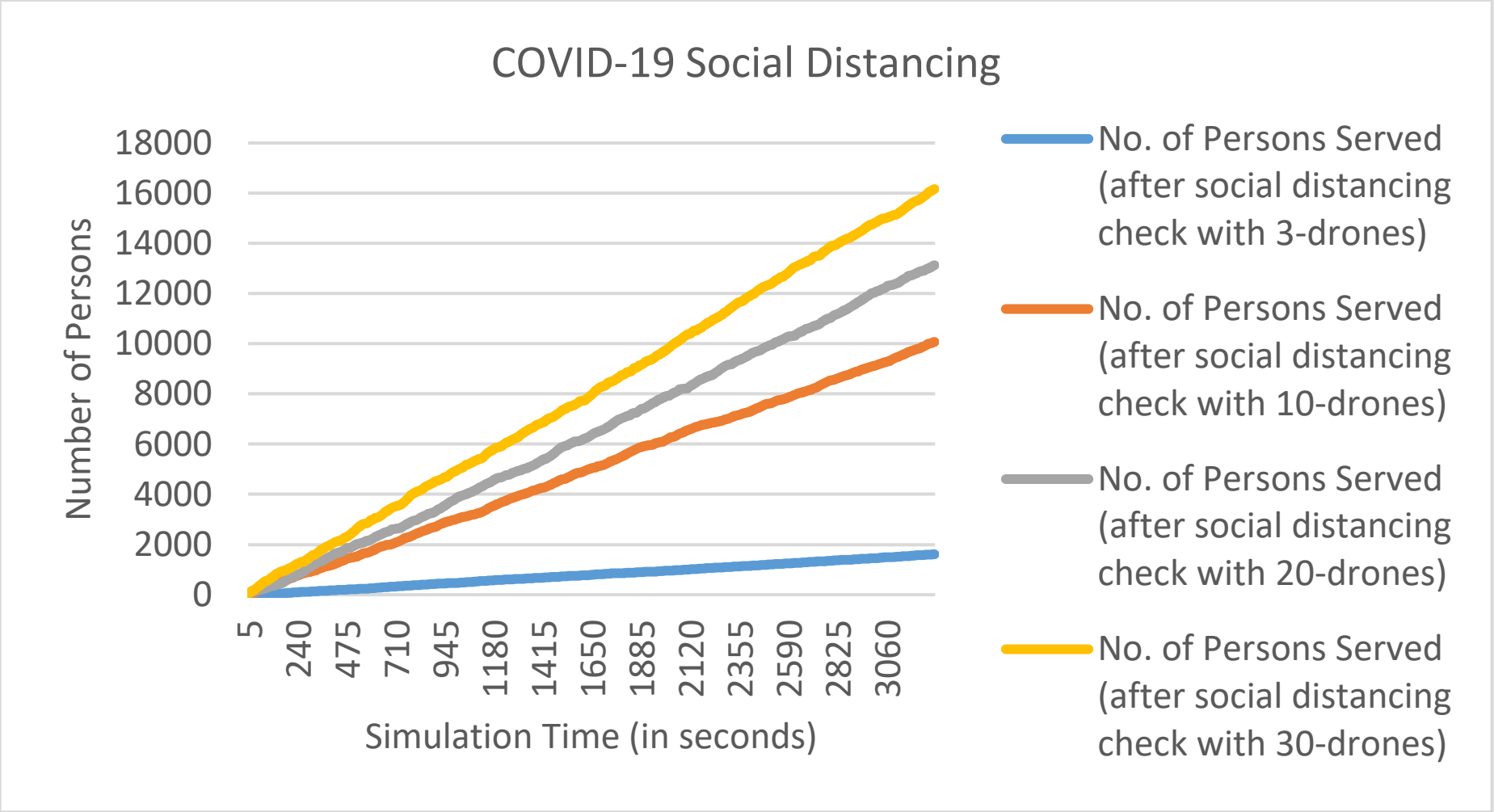

**Figure 22:** Comparative analysis of the number of persons served with medicine supply after social distancing check (the maximum number of parallel medicine supply units=10, the maximum time taken to supply a medicine= 120 seconds)

### *5.5 Case 5: Drone-based System and Simulation for Police Monitoring and COVID Control Room*

Fig. 23 and fig. 24 show the results that can be displayed in the control room in addition to the statistics shown in other cases. These statistics are simulated using AnyLogic simulator. Fig. 23 shows the variations in drones used over the number of simulation days in the sanitization process. This statistic is variable and it varies with simulation. However, Fig. 23 shows a static view of statistics between 0 to 600 days. Here, it is observed that 120 to 356 drones (cumulative) are used for the sanitization process. The cumulative drone number over the simulation time (in days) shows the drone usage variations. In this experimentation, the same drone is counted twice if it is re-used for sanitization. Fig. 24 shows the percentage of drone utilization. Results show that the percentage of drone utilization varies from 0% to 80% approximately. These statistics are variable as well and it varies with simulation. Fig. 24 shows the results for 30-drones. These statistics can be helpful to control room officials for analyzing the drone situation and instructions could be passed accordingly. If the utilization is higher than the chance of drone to deplete its battery, at a much faster rate, is higher as well. Thus, the minimum utilized drones can be used for subsequent operations.

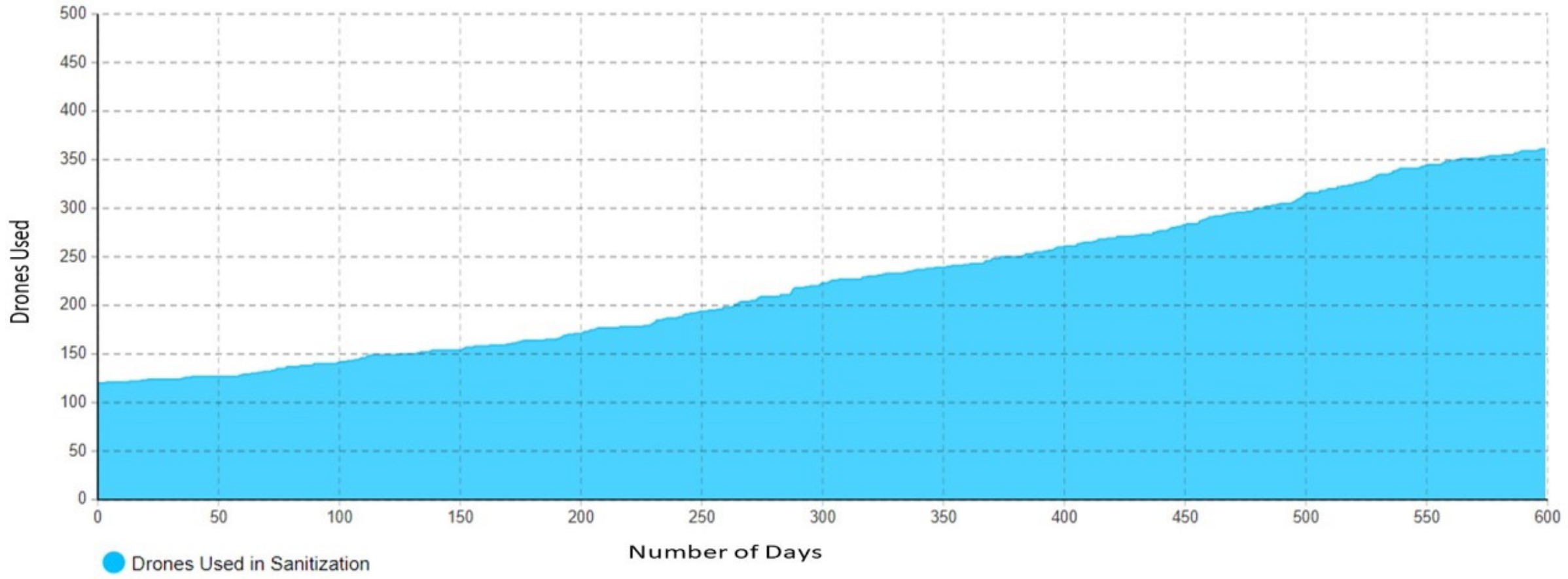

**Figure 23:** Cumulative Drone utilization over the simulation time

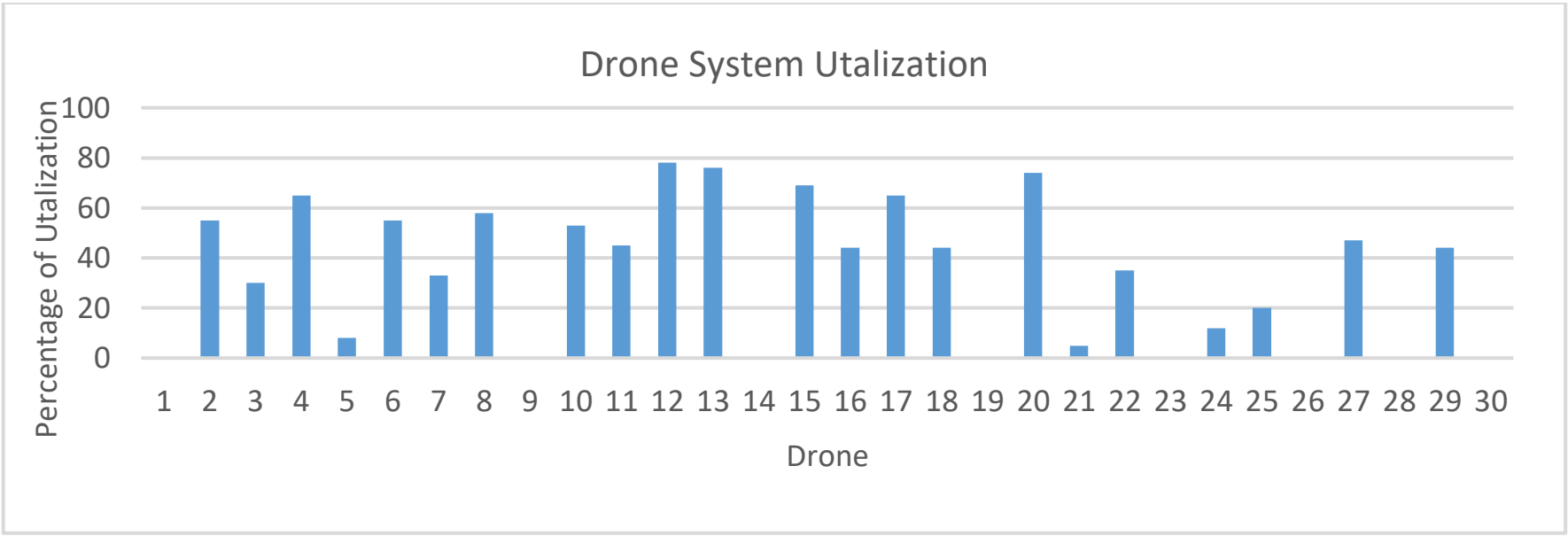

**Figure 24:** Percentage of drone-system utilization

## 5.6 *Case 6: Drone-based Simulation for Performance Analysis in COVID-19 Scenario*

**Simulation Model**: Fig. 25 shows the drone-based sanitization system designed and programmed for simulating the sanitization process in the COVID-19 system with a sequential drone-movement strategy. This model is simulated using the JaamSim simulator.

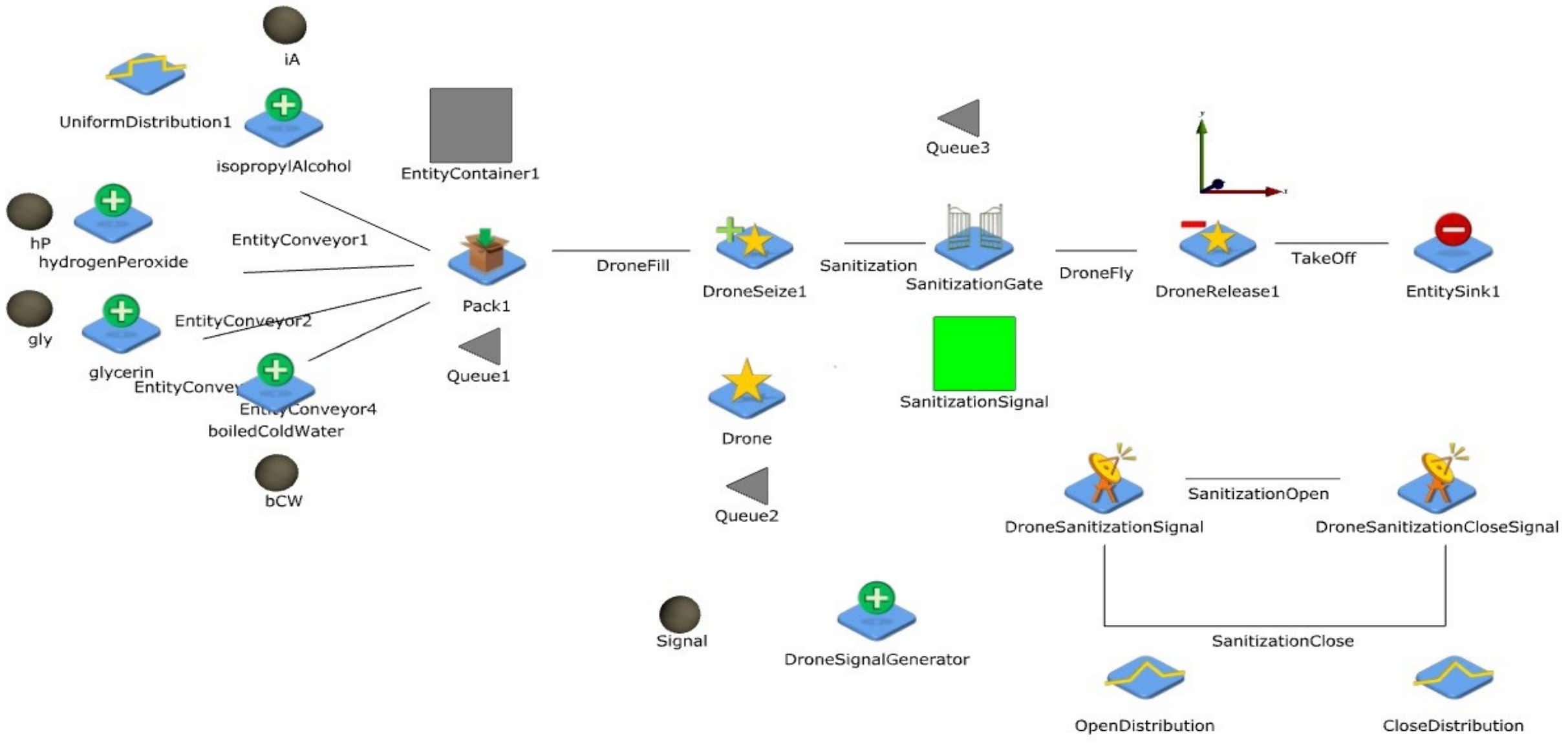

**Figure 25:** Drone-based Sanitization System with sequential drones movement

**Assumptions:** It is assumed that the maximum time taken to send a signal to the drone is 10 seconds. Here, triangular distribution (compared to a normal distribution) is used for setting the signal time variation in the simulation model because distribution is expected to be skewed with completion of the area under observation at the minimum, maximum and modal values.

**Outcomes & Discussions**: This sub-section explains the various parameters used for performance measurement. These are explained as follows.

*Ground to Drones Signal Transmission Time Analysis:* Fig. 26 shows the variations in signal transmission time from a ground transmitter to the drone's receiver. This variation is observed for an infinite period. Fig. 26 shows the variations for the past one hour (0 to 3600 seconds). Here, "-" sign is an indication of past time only. Results show that signal duration varies from 0 (minimum) to 9 seconds (maximum). Mean and standard deviation values of time variation are found to be 4.1 seconds and 3.7 seconds for 10 hours.

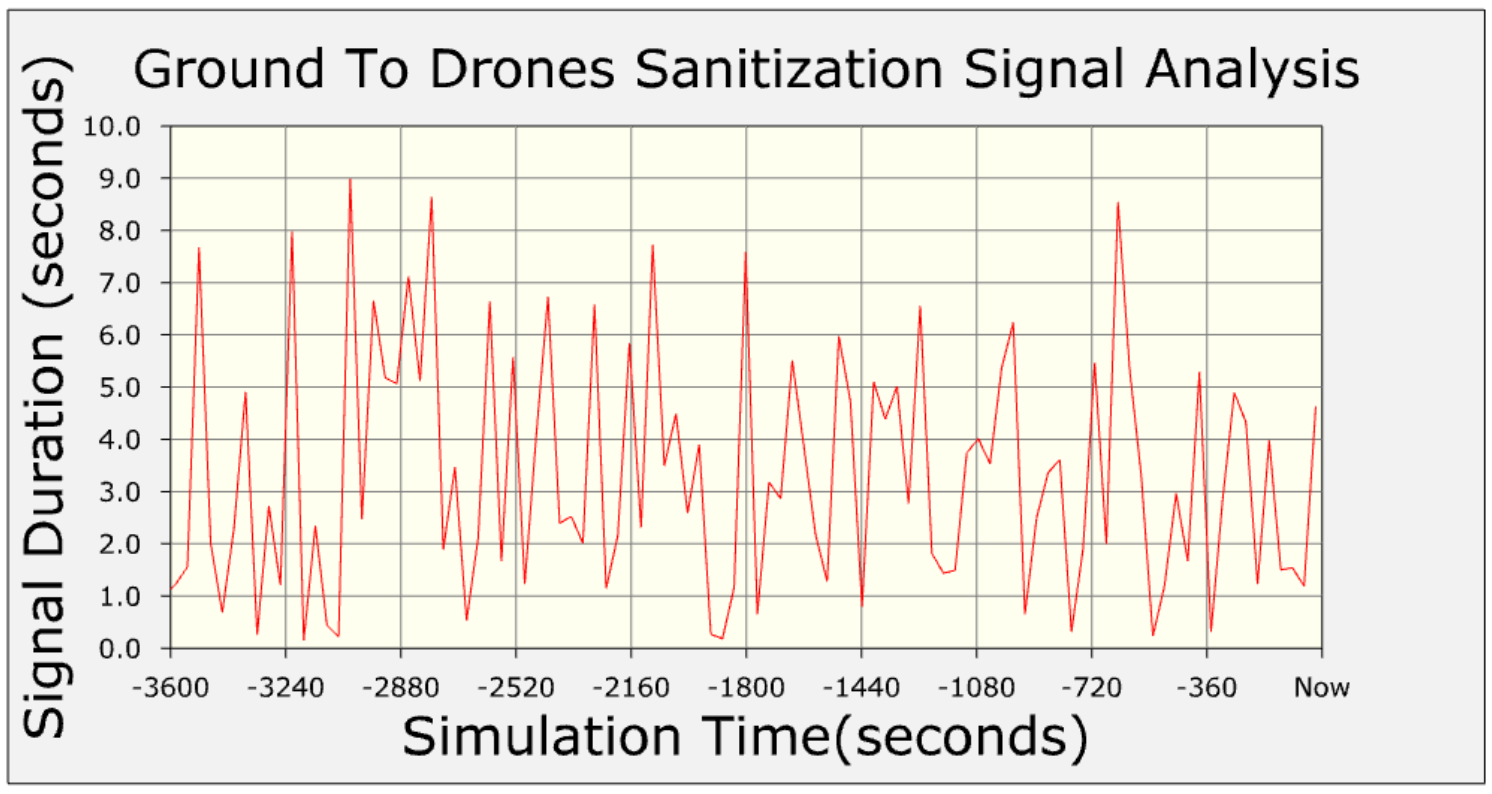

**Figure 26:** Ground to Drones Signal Transmission Time Analysis

*Number of Drones Used*: Fig. 27 shows the number of drones used over simulation time. In the proposed simulation model, it is observed that a minimum of 25 and a maximum of 55 drones are used at a time. This usage includes drones while operating inside the zone, filling the sanitizer, flying with no operation, and landing operations.

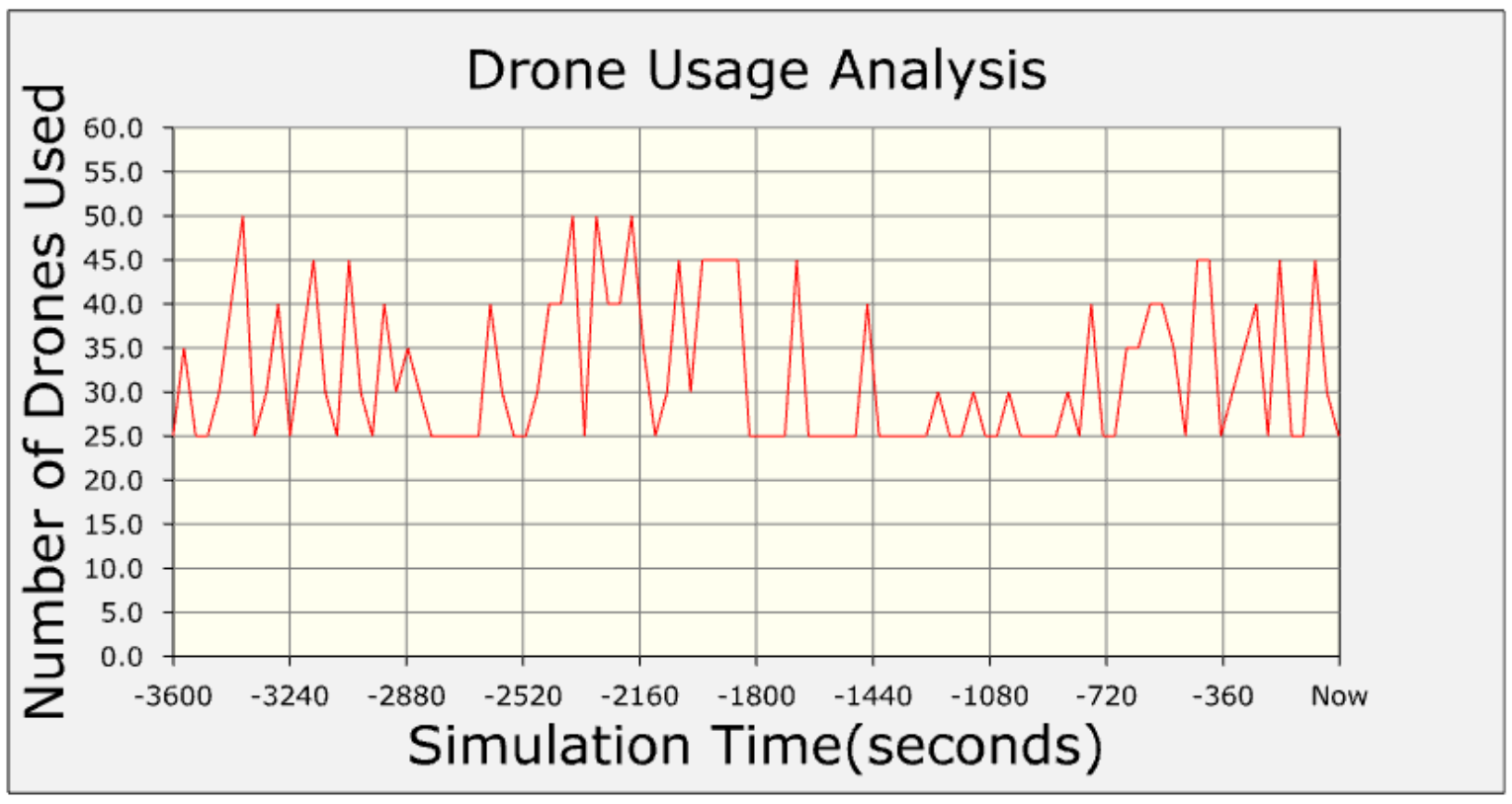

**Figure 27:** Drone Usage Analysis

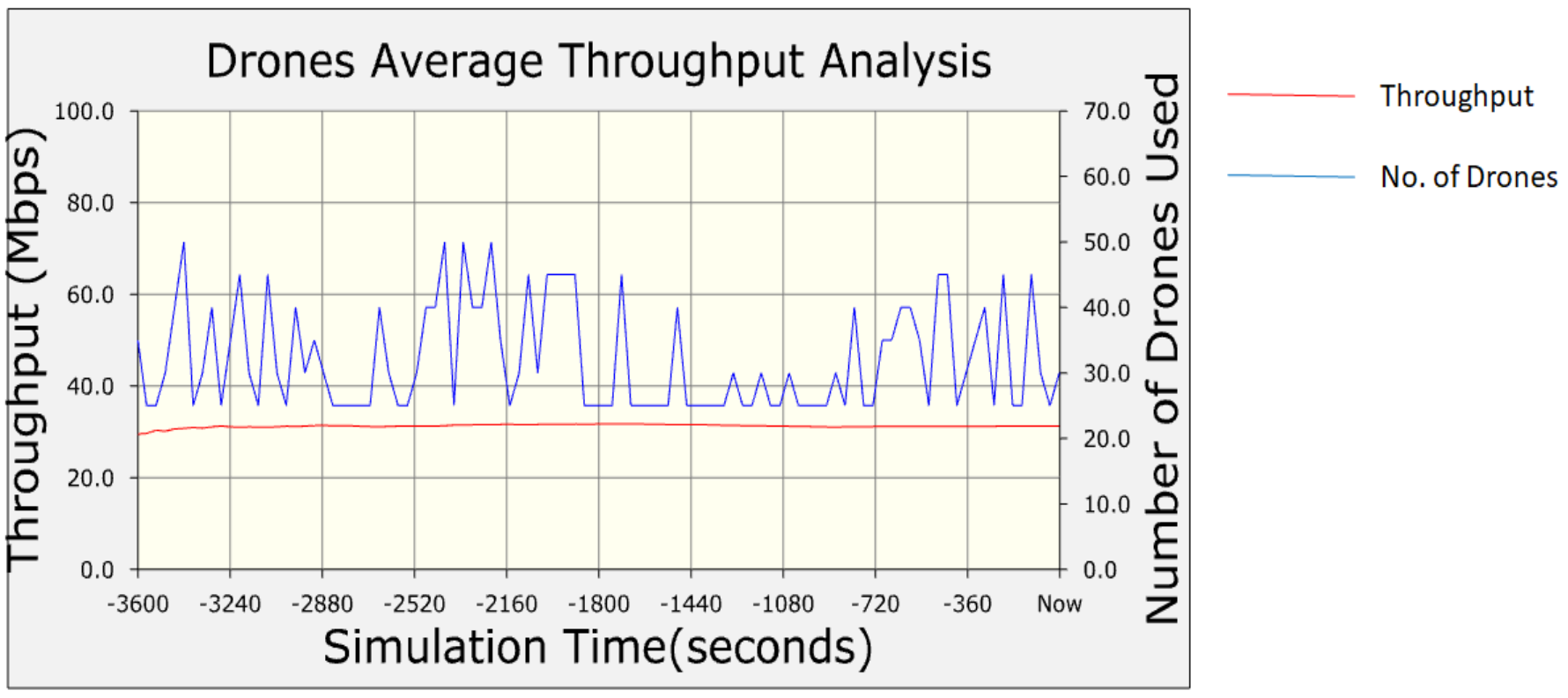

**Figure 28:** Drone Average Throughput Analysis with Variations in Number of Drones

Fig. 28 shows the average throughput variation analysis. Equation (i) is used to compute the throughput [33]. Here, Bit Error Ratio (BER) is the ratio of the number of bits errors by the total number of transmitted bits during the total transmission time. With an average number of drones usage around 22, average throughput varies from 35 Mbps (minimum) to 70 Mbps (maximum).

$$Throughput = \frac{256*8*N_{success}*(1-BER)}{Total\ Packet\ Transmission\ Time} \quad \text{(i)}$$

Fig. 29 shows a drone-based sanitization system with parallel drone movements. In this experimentation, it is realized that the average throughput lies between 35 Mbps to 80 Mbps with an average number of drones' usage around 17.

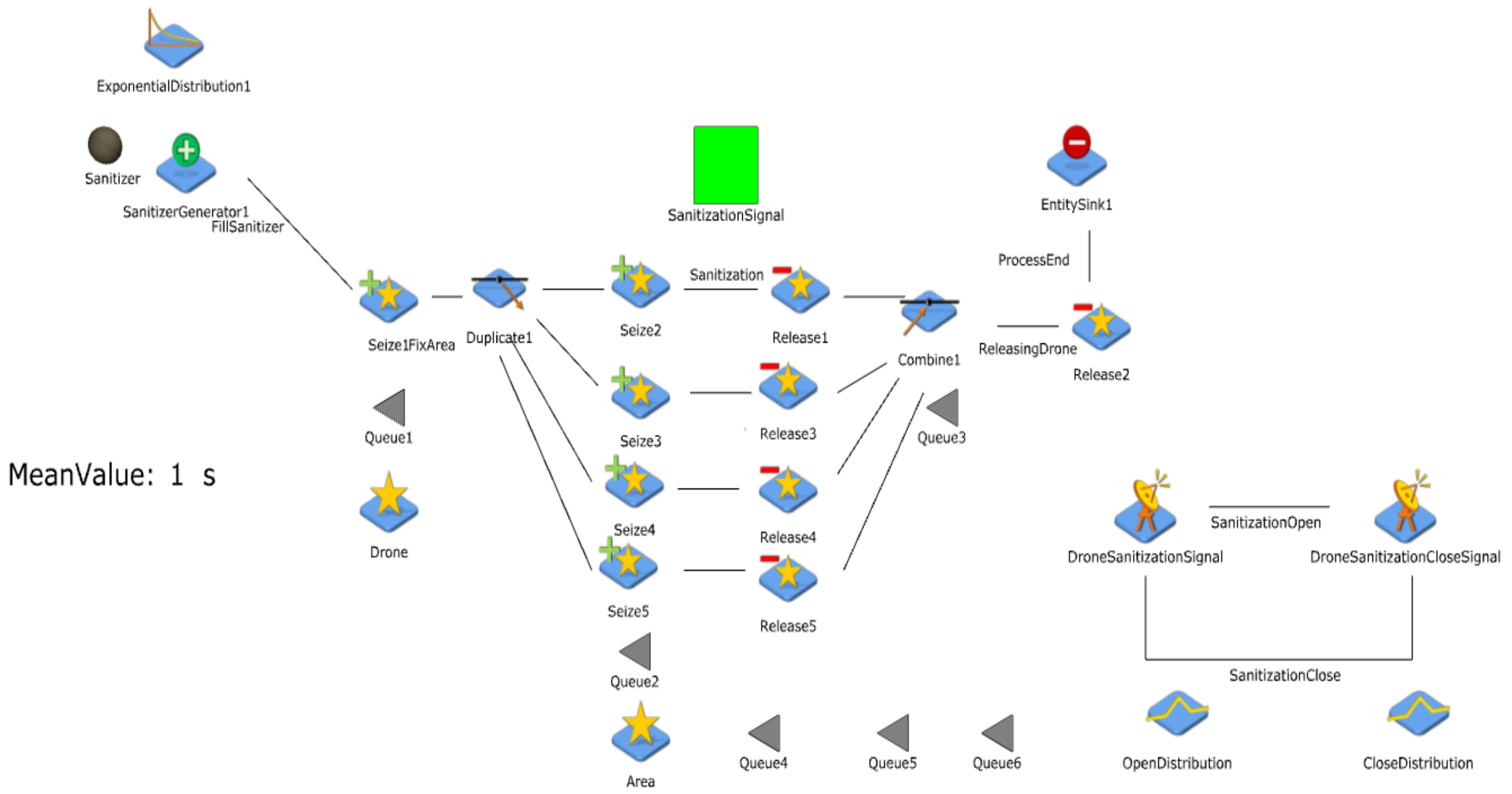

**Figure 29:** Drone-based sanitization system with parallel drones movement

Fig. 30 shows the comparative analysis of drone movement and zone transfer strategy with the existing approach [34] with simulation time variation. Nageli et al. [34] use video graphing-based multi drone movement and collision avoidance algorithms. Results show that Nageli et al. algorithm takes more time in execution as compared to proposed single-layer algorithms such as fixed-area (algorithm 2), zig-zag (algorithm 3), and parallel (algorithm 5) in the majority of the cases. The two-layered zone transfer approach (algorithm 6) is expensive in some cases.

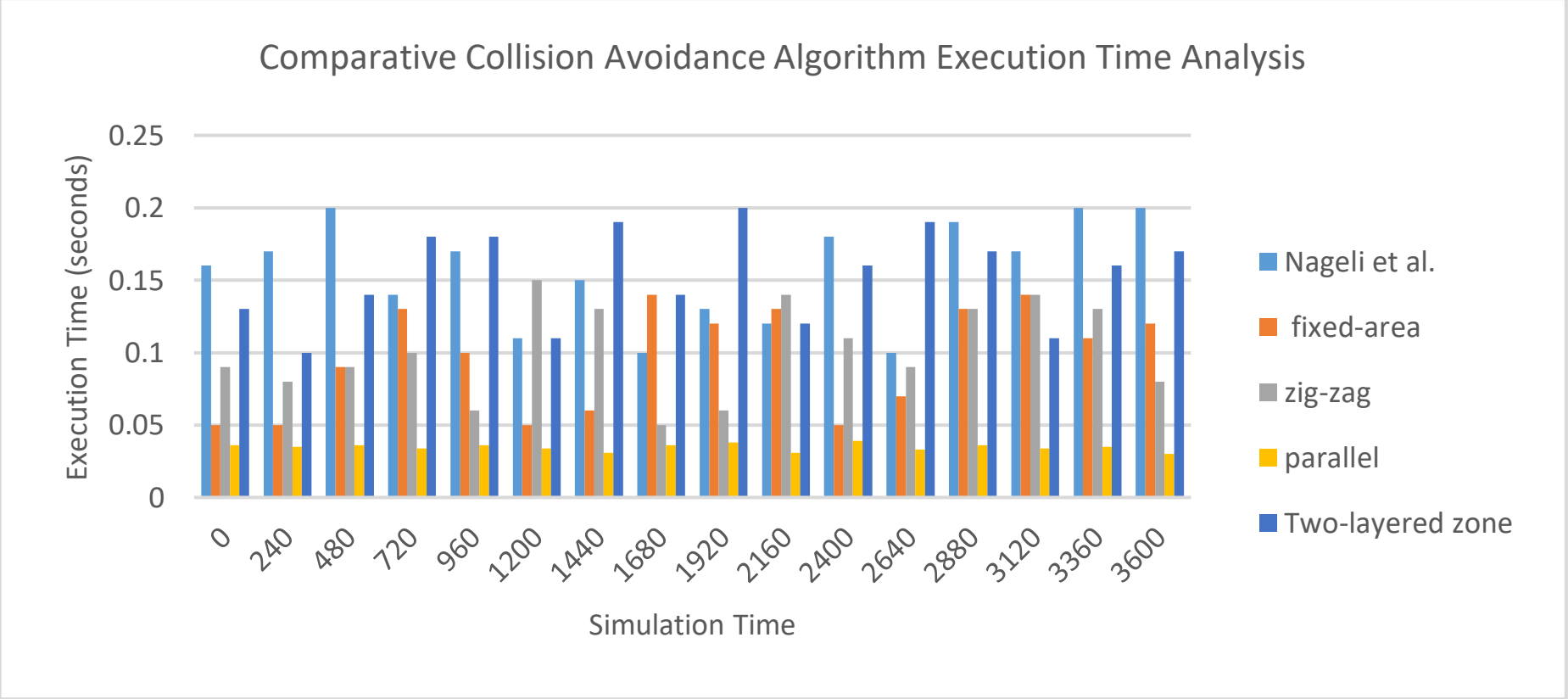

**Figure 30:** Comparative Collision Avoidance Algorithm Execution Time Analysis

## 6. Conclusions and Future Directions

This paper proposed a UAV-based smart healthcare system for COVID-19 monitoring, sanitization, social distancing, data analysis, and statistics generation for the control room. Our framework gathers data by either through wearable sensors, movement sensors deployed in the targeted areas, or through thermal image processing. The data is processed through a multilayered architecture for analysis and decision-making. In multilayered architecture, edge computing controls the proposed drones' collision-resistant strategies. Whereas, fog and cloud computing techniques build commuters and patient profiles before making decisions. The proposed approach is demonstrated with implementation and simulation. In an implementation, it is observed that a large distance can be covered within a short period and the proposed drone-based healthcare system is effective for COVID-19 operations in Delhi/NCR regions. In the simulation, the proposed approach is tested for indoor and outdoor activities. Results show that a distance of 1200 kilometers can be covered in 2293 to 18900 minutes with a variation of 3 to 30 drones. In an indoor activity, thermal image-based patient identification is found to be very effective for COVID-19 pandemic.

The simulation studies (Case-2 to Case-6) in the proposed drone-based smart healthcare system have limitations such as it considers the movement of drones in an ideal scenario. In the real-scenario, environmental conditions affects its movement and can change the statistics of its usage. Further, it is assumed that the drone used for indoor imaging and sanitization is compatible with its operations. In the real-scenario, a compact-drone is required for similar operations. Thus, compact-drone design aspects should be considered to analyze the real-facts.

## *6.1 Open Challenges and Future Directions*

Although the proposed drone-based networked system and methods facilitate in combating COVID-19, it can be further enhanced in a large scope keeping the following aspects into consideration.

1. *Large scale medicine delivery*: In recent times, chloroquine and hydroxychloroquine is delivered using drone-based system to fight against COVID-19. The COVID-19 medicines are under trial. In the future, the feasibility of large-scale medicine delivery with different collision-resistant strategies can be explored keeping infrastructure constraints in consideration.
2. *Residents' records and scanning*: Resident record is another important parameter to ensure that every person living in an area is scanned. Although resident welfare association keeps the records of every resident but many people live outside resident societies as well. Thus, it is important to keep the record of everyone and match with scanned population. This way, we can ensure that there are no resident with initial symptoms of COVID-19.
3. *Scanning services in case of lack of medical infrastructure*: It has been observed that many doctors were reluctant to handle patients when COVID-19 pandemic started. This is because of lack of medicine facilities, patient behaviours, and long distance travel to provide services. In such cases, non-medical person capable of flying drones to facilitate COVID-19 scanning or medicine supply can help and speed up the scanning and testing processes.
4. *Normal (Manual) vs drone-based thermal scanning statistics*: The comparative analysis of manual vs drone-based thermal scanning is required to show the importance of both types of systems in medical system. For example, normal (manual) system is preferred if all medical facilities (including personal protection equipment, a surgical gown, gloves, respiratory protection, eye protection, face shield etc.) are timely available. It has been observed that drone-based system is preferred in case of scarcity of medical facilities (like at initial stages of COVID-19 pandemic).
5. *Large deployment of mini-drones for indoor operations*: We have used drones to perform thermal scanning in both real-time and simulation-based experimentations. In both cases, residential areas are considered for scanning. Here, drones are used to scan people in both high-rise buildings and pedestrians in real-time experimentation. Simulation considered the scanning of pedestrians only. However, multiple mini-drones flying machines are required to perform indoor scanning operation in case people do not take self-initiatives to come to balconies for scanning.
6. *Smart mini-drones and accessibility in internal boundaries of the home/hospital*: There is a need to test the drones capable to operate at a long distance with indoor operations. The accessibility to operate such drones and data collection without interruption is a major issues that need to be tackled in future.
7. *Efficient framework implementation*: We used different subsystems of the proposed system with temporary arrangements with major concentration over combating COVID-19 pandemic and its cases. For example, a group of machine at scanning sites were used as edge network and computing system for filtering the required data and sharing with hospital under federal government monitoring and control. However, there is a need to standardize such computing platforms and their efficiencies before actual use. Thus, sub-system services including the use of computing infrastructure, security primitives and protocols, and long term secure big data storage will be implemented.
8. *Fool-Proof COVID-19 system and comparative analysis*: In the proposed system, the comparative analysis with state-of-the art work is limited to collision avoidance algorithms. However, there is a need to perform more such comparative analysis like performance analysis of proposed system with other pandemic and medical systems, and comparative security analysis of medical systems.
9. *Large scale medical infrastructure integration*: There is a need to develop an integrated medical system for fast monitoring and collective support system that is capable of collecting long distant patient data and provide services at large scale. Such systems can consider environment conditions and other barriers in flying drones, collecting data, and maintain the data security as per government policies.

**Acknowledgments**

We thank Prof. Peter Sloot (Editor-in-Chief) and anonymous reviewers for their valuable comments and suggestions for improving our paper.

## References

[1] L. Li, Q. Zhang, X. Wang, J. Zhang, T. Wang, T. L. Gao, W. Duan, K. K. fai Tsoi, and F. Y. Wang, "Characterizing the Propagation of Situational Information in Social Media During COVID-19 Epidemic: A Case Study on Weibo," IEEE Transactions on Computational Social Systems, vol. 7, no. 2, pp. 556–562, 2020.

[2] C. Hopkins and N. Kumar, "Loss of sense of smell as marker of COVID-19 infection," Ent Uk, pp. 19–20, 2020.

[3] J. T. Wu, K. Leung, and G. M. Leung, "Nowcasting and forecasting the potential domestic and international spread of the 2019-nCoV outbreak originating in Wuhan, China: a modelling study," The Lancet, vol. 395, no. 10225, pp. 689–697, 2020.

[4] "Draganfly selected to globally integrate breakthrough health diagnosis technology immediately onto autonomous camera's and specialized drones to combat coronavirus (covid19) and future health emergencies." [Online]. Available: https://apnews.com/Globe%20Newswire/dc01344350423d7d64c99ebbe8fb7548

[5] "Drones and the coronavirus: How drones are helping containment efforts." [Online]. Available: https://uavcoach.com/drones-coronavirus/

[6] "Corona combat drone: Indian robotics solution launches corona combat drone to fight covid-19, government news, et government." [Online]. Available: https://government.economictimes.indiatimes.com/news/technology/indianrobotics-solution-launches-corona-combat-drone-to-fightcovid-19/75077517

[7] "Drone technology: A new ally in the fight against COVID-19." https://www.mdlinx.com/internal-medicine/article/6767 (accessed Apr. 15, 2020).

[8] "Drone Association of Kerala: Latest News & Videos, Photos about Drone Association of Kerala | The Economic Times." https://economictimes.indiatimes.com/topic/Drone-Association-of-Kerala (accessed Apr. 15, 2020).

[9] "Covid-19 lockdown: Authorities rely on drone eye to maintain vigil - The Economic Times." https://economictimes.indiatimes.com/news/politics-and-nation/covid-19-lockdown-authorities-rely-on-drone-eye-to-maintain-vigil/articleshow/75112745.cms (accessed Apr. 15, 2020).

[10] M. J. Lum, J. Rosen, H. King, D. C. Friedman, G. Donlin, G. Sankaranarayanan, B. Harnett, L. Huffman, C. Doarn, T. Broderick, and B. Hannaford, "Telesurgery via unmanned aerial vehicle (UAV) with a field deployable surgical robot," Studies in Health Technology and Informatics, vol. 125, pp. 313–315, 2007.

[11] D. Cˆamara, "Cavalry to the rescue: Drones fleet to help rescuers operations over disasters scenarios," 2014 IEEE Conference on Antenna Measurements and Applications, CAMA 2014, pp. 1–4, 2014.

[12] S. J. Kim, G. J. Lim, J. Cho, and M. J. Cote, "Drone-Aided Healthcare Services for Patients with Chronic Diseases in Rural Areas," Journal of Intelligent and Robotic Systems: Theory and Applications, vol. 88, no. 1, pp. 163–180, 2017.

[13] R. F. Graboyes and B. Skorup, "Medical Drones in the United States and a Survey of Technical and Policy Challenges," SSRN Electronic Journal, no. February, 2020

[14] C. F. Peng, J. W. Hsieh, S. W. Leu, and C. H. Chuang, "Drone-based vacant parking space detection," Proceedings - 32nd IEEE International Conference on Advanced Information Networking and Applications Workshops, WAINA 2018, vol. 2018-January, pp. 618–622, 2018.

[15] S. Pirbhulal, W. Wu, G. Li, and A. K. Sangaiah, "Medical Information Security for Wearable Body Sensor Networks in Smart Healthcare," IEEE Consumer Electronics Magazine, vol. 8, no. 5, pp. 37–41, 2019.

[16] H. Ullah, N. Gopalakrishnan Nair, A. Moore, C. Nugent, P. Muschamp, and M. Cuevas, "5G communication: An overview of vehicle-to-everything, drones, and health care use cases," IEEE Access, vol. 7, pp. 37251–37268, 2019.

[17] R. W. Jones and G. Despotou, "Unmanned aerial systems and healthcare: Possibilities and challenges," in 2019 14th IEEE Conference on Industrial Electronics and Applications (ICIEA), June 2019, pp. 189–194.

[18] A. Islam and S. Young Shin, "A blockchain-based secure healthcare scheme with the assistance of unmanned aerial vehicle in Internet of Things," Computers and Electrical Engineering, vol. 84, p. 106627, 2020.

[19] S. C. Sethuraman, V. Vijayakumar, and S. Walczak, "Cyber Attacks on Healthcare Devices Using Unmanned Aerial Vehicles," Journal of Medical Systems, vol. 44, no. 1, 2020.

[20] K. Lee and J. Park, "Application of geospatial information of neighborhood park for healthcare of local residents," Journal of Medical Imaging and Health Informatics, vol. 7, pp. 674– 679, 06 2017.

[21] S. Ullah, K. I. Kim, K. H. Kim, M. Imran, P. Khan, E. Tovar, and F. Ali, "UAV-enabled healthcare architecture: Issues and challenges," Future Generation Computer Systems, vol. 97, pp. 425–432, 2019. [Online]. Available: https://doi.org/10.1016/j.future.2019.01.028

[22] B. Skorup and C. Haaland, "How Drones Can Help Fight the Coronavirus," SSRN Electronic Journal, 2020.

[23] B. M. Harnett, C. R. Doarn, J. Rosen, B. Hannaford, and T. J. Broderick, "Evaluation of unmanned airborne vehicles and mobile robotic telesurgery in an extreme environment," Telemedicine and e-Health, vol. 14, no. 6, pp. 539–544, 2008.

[24] C. Todd, M. Watfa, Y. El Mouden, S. Sahir, A. Ali, A. Niavarani, A. Lutfi, A. Copiaco, V. Agarwal, K. Afsari, C. Johnathon, O. Okafor, and M. Ayad, "A proposed UAV for indoor patient care." Technology and health care: official journal of the European Society for Engineering and Medicine, vol. 1, pp. 1–8, 2015.

[25] C. A. Thiels, J. M. Aho, S. P. Zietlow, and D. H. Jenkins, “Use of unmanned aerial vehicles for medical product transport,” Air Medical Journal, vol. 34, no. 2, pp. 104–108, 2015.
[26] “How fast can drones fly? max speed of drones - 3d insider.” [Online]. Available: https://3dinsider.com/drone-speed/
[27] M. Chiang and T. Zhang, “Fog and iot: An overview of research opportunities,” IEEE Internet of Things Journal, vol. 3, no. 6, pp. 854–864, Dec 2016.
[28] N. Wickramasinghe and F, Bodendorf, “Delivering superior health and wellness management with iot and analytics.”, Springer International Publishing, 2020.
[29] D. Palossi, A. Loquercio, F. Conti, E. Flamand, D. Scaramuzza, and L. Benini, “A 64-mW DNN-Based Visual Navigation Engine for Autonomous Nano-Drones,” IEEE Internet of Things Journal, vol. 6, no. 5, pp. 8357–8371, 2019.
[30] N. Hossein Motlagh, T. Taleb, and O. Arouk, “Low-Altitude Unmanned Aerial Vehicles-Based Internet of Things Services: Comprehensive Survey and Future Perspectives,” IEEE Internet of Things Journal, vol. 3, no. 6, pp. 899–922, 2016.
[31] Anylogic Simulator, https://www.anylogic.com/ [Last Accessed on: May 03, 2020]
[32] JaamSim Simulator, https://jaamsim.com/ [Last Accessed on: May 03, 2020]
[33] Shi, Y., Enami, R., Wensowitch, J. and Camp, J., 2018, April. Measurement-based characterization of LOS and NLOS drone-to-ground channels. In *2018 IEEE Wireless Communications and Networking Conference (WCNC)* (pp. 1-6). IEEE.
[34] Nägeli, T., Meier, L., Domahidi, A., Alonso-Mora, J. and Hilliges, O., 2017. Real-time planning for automated multi-view drone cinematography. *ACM Transactions on Graphics (TOG)*, *36*(4), pp.1-10.
[35] Shreshth Tuli, Shikhar Tuli, Rakesh Tuli and Sukhpal Singh Gill, “Predicting the Growth and Trend of COVID-19 Pandemic using Machine Learning and Cloud Computing” *Internet of Things, Elsevier*, vol. 11, 2020. https://doi.org/10.1016/j.iot.2020.100222
[36] J. Gubbi, R. Buyya, S. Marusic, and M. Palaniswami, Internet of Things (IoT): A Vision, Architectural Elements, and Future Directions, Future Generation Computer Systems, 29(7): 1645-1660, Elsevier Science, Amsterdam, The Netherlands, September 2013.
[37] Pereira, Adrián Arenal, Jordán Pascual Espada, Rubén González Crespo, and Sergio Ríos Aguilar. "Platform for controlling and getting data from network connected drones in indoor environments." Future Generation Computer Systems 92 (2019): 656-662.
[38] Al-Sa’d, Mohammad F., Abdulla Al-Ali, Amr Mohamed, Tamer Khattab, and Aiman Erbad. "RF-based drone detection and identification using deep learning approaches: An initiative towards a large open source drone database." Future Generation Computer Systems 100 (2019): 86-97.
[39] Li, Yujie, Huimin Lu, Yoshiki Nakayama, Hyoungseop Kim, and Seiichi Serikawa. "Automatic road detection system for an air–land amphibious car drone." Future Generation Computer Systems 85 (2018): 51-59.
[40] De Benedetti, M., D'Urso, F., Fortino, G., Messina, F., Pappalardo, G. and Santoro, C., 2017. A fault-tolerant self-organizing flocking approach for UAV aerial survey. *Journal of Network and Computer Applications*, *96*, pp.14-30.
[41] Pace, P., Aloi, G., Caliciuri, G. and Fortino, G., 2016. A mission-oriented coordination framework for teams of mobile aerial and terrestrial smart objects. *Mobile Networks and Applications*, *21*(4), pp.708-725.
[42] M. Whaiduzzaman, M. Hossain, A. R. Shovon, S. Roy, A. Laszka, R. Buyya, and A. Barros, 2020. A Privacy-preserving Mobile and Fog Computing Framework to Trace and Prevent COVID-19 Community Transmission. *arXiv preprint arXiv:2006.13364*.